\documentclass[twocolumn,showkeys,showpacs,amsmath,amssymb,superscriptaddress,nofootinbib]{revtex4}
\usepackage{amssymb}
\usepackage{amsmath}
\usepackage{graphicx}
\normalbaselines
\begin{document}

\title{Polarization and stratification of  axionically active plasma \\ in a dyon magnetosphere}

\author{Alexander B. Balakin}
\email{Alexander.Balakin@kpfu.ru}
\author{Dmitry E. Groshev }
\email{groshevdmitri@mail.ru}
\affiliation{Department of General Relativity and
Gravitation, Institute of Physics,Kazan Federal University, Kremlevskaya street 18, 420008, Kazan,
Russia}

\begin{abstract}
The state of a static spherically symmetric relativistic axionically active  multi-component plasma in the gravitational, magnetic and electric fields of an axionic dyon is studied in the framework of the Einstein - Maxwell - Fermi - Dirac - axion theory. We assume that the equations of axion electrodynamics, the covariant relativistic kinetic equations, and the equation for the axion field with modified periodic and Higgs-type potentials are nonlinearly coupled; the gravitational field in the dyon exterior is assumed to be fixed and to be of the Reissner-Nordstr\"om type. We introduce the extended Lorentz force, which acts on the particles in the axionically active plasma, and analyze the consequences of this generalization. The analysis of exact solutions, obtained in the framework of this model for the relativistic Boltzmann electron-ion and electron-positron plasmas, as well as, for degenerated zero-temperature electron gas, shows that the phenomena of polarization and stratification can appear in plasma, attracting attention to the axionic analog of the known Pannekoek-Rosseland effect.
\end{abstract}



\maketitle

\section{Introduction}

The effect of electric polarization of the equilibrium multi-component plasma in the external gravitational field has been known since the  1920s due to the classical works  of Pannekoek \cite{P} and Rosseland \cite{R}. This effect was described, first, in the context of electric polarization of the non-relativistic electron-ion isothermal plasma in the external Newtonian gravitational field.  Since the rest mass of the ion exceeds the electron mass, the profile of the ion density in the gravitational field, designed with respect to altitude, differs from the profile of the electron density, thus violating the local electric neutrality of the equilibrium isothermal plasma. As a consequence, the compensating electric field, known as the Pannekoek - Rosseland electric field, appears in the plasma. This model became very convenient for various applications to Earth's, solar and pulsar plasmaphysics in non-relativistic and relativistic versions, for weak and strong gravitational fields (see, e.g., \cite{1,2,3,4,5,6} and references therein).

All the known calculations of the Pannekoek - Rosseland electric field in plasma are based on the Faraday - Maxwell version of electrodynamics. Nowadays, there exists serious interest in astrophysical applications of its extension, the axion electrodynamics, which takes into consideration  massive pseudo-Goldstone bosons \cite{PQ,Weinberg,Wilczek}, interacting with photons.
These light bosons (axions in modern terminology) are the most probable candidates for dark matter particles \cite{add1,ADM1,ADM2,ADM3,ADM4,ADM41,ADM5,ADM6,ADM7,add3,add4,ADM8}, and their contribution into the Universe energy balance is estimated to be about 23\%. There are two constitutive elements in the axion electrodynamics: first, the electromagnetic field with the vector potential $A_k$, second, the pseudoscalar field $\phi$. The simplest Lagrangian of interaction between the pseudoscalar field and photons was introduced in \cite{Ni77}; its modifications and new applications have been considered in many works (see, e.g., \cite{Sikivie83,Wilczek87,BG,Symm}). In the works \cite{AAP1,AAP2,AAP3} various aspects of the theory of the axionically active relativistic plasma were studied; in particular, the propagation of transversal electromagnetic waves with subluminal phase velocity, which can not exist in the standard plasma, but can appear due to the axion-photon coupling, was described.

Monopoles in the axion environment become dyons due to the Witten effect \cite{Witten}; the influence of this effect on the dyon-axion dynamics was studied in \cite{FP1983}; further applications and history of investigations of the axion-monopole coupling in the QCD context are discussed in \cite{dyon2016}.  Wilczek \cite{Wilczek87} has discussed the idea of axionic dyon in the context of axion electrodynamics. In such an object the pseudoscalar (axion) field produces the radial electric field in the presence of the radial magnetic field. Later this model was studied in the framework of the full-format Einstein-Maxwell-axion theory in the context of description of horizons of axionic black holes \cite{Y}. In \cite{BZ17} it was shown that the axionic dyon can possess a regular electric field equal to zero at the center. The next step is to use the model of axionic dyon as a "theoretical laboratory" for modeling of physical processes in the exterior of this object, under the assumption that the axionic dyon can possess  strong gravitational, magnetic and electric fields.

In this paper we consider the polarization of relativistic plasma taking into account the axionic analog of the Pannekoek - Rosseland effect in the plasma surrounding the axionic dyon. The idea is the following. Let us imagine the magnetic monopole with strong radial magnetic field, pseudoscalar (axion) field and relativistic multi-component plasma in the object exterior. The axion-photon interactions produce the radial electric field \cite{Wilczek87}, thus inducing the plasma polarization of the first type. Also, the multi-component plasma is polarized in the strong gravitational field of the monopole; it is the polarization of the second type. We use the term {\it axionically active} plasma in analogy with the well-known term magneto-active plasma. In the magneto-active plasma the interaction of charged particles with the magnetic field is known to lead to various sophisticated phenomena. In the axionically active plasma the axion-photon coupling is assumed to be the source of new effects in plasma.

A number of questions arise, when we consider the physical sense of such model.
The first series of questions concerns the profile of the resulting electric field in the polarized plasma. Is the mutual compensation of the electric fields of the first and second types possible? Does the plasma stratification, resulting from this polarization, exist? What is the typical size of the electric stratum in the polarized plasma? Whether the backreaction of the plasma polarization on the pseudoscalar field plays an essential role in the formation of the axion density profile in the vicinity of the dyon?

The second series of questions is focused around the polarization of {\it electron-positron} plasma, which can surround the object with a strong magnetic field. The classical Pannekoek - Rosseland effect in the electron-positron plasma is absent, since the masses of the electrons and positrons coincide. However, the dyon possesses a radial electric field induced due to the axion-photon coupling, thus the electron-positron plasma has to be inevitably polarized. One of the questions arising in this context is about the plasma reaction: can the cumulative effect lead to a self-compensation of the electric field, or, on the contrary, to the plasma stratification?

The third series of questions is connected with the problem of axionic modification of the Lorentz force. The Lagrangian describing the axion electrodynamics contains two invariant terms: $\frac14 F_{mn}F^{mn}$ and $\frac14 \phi F^{*}_{mn}F^{mn}$ (see \cite{Ni77}); the first one is constructed using the Maxwell tensor $F_{mn}$, while the second term includes the dual Maxwell tensor (pseudotensor) $F^{*}_{mn}$, multiplied by the pseudoscalar $\phi$.  We formulate the hypothesis that the force, which acts on the plasma particles in the electromagnetic and axion fields (the {\it axionically extended} Lorentz force) can be represented using the linear combination $F_{mn}+ \nu \phi F^{*}_{mn}$ instead of pure $F_{mn}$. The parameter $\nu$ is equal to zero, if the true force is the pure Lorentzian one, and $\nu=1$, if there exists an equivalence between these two terms, $F_{mn}$ and $\phi F^{*}_{mn}$. We make the calculations for both cases $\nu=0$ and $\nu=1$, keeping in mind that the predictions of theory could be tested, and the true value of the parameter $\nu$ be estimated from astrophysical observations.

We analyze the state of the static spherically symmetric equilibrium plasma configuration in the framework of Einstein - Maxwell - Fermi - Dirac - axion theory. This means the following: the gravitational field in the dyon exterior is presented by the Reissner-Nordstr\"om solution to the Einstein equations; the electric and magnetic fields are the solutions
to the equations of axion electrodynamics with the sources generated by the plasma and axion field; the distribution functions for the components of plasma are the solutions to the relativistic covariant kinetic  equations; the pseudoscalar (axion) field is described by the solution to the master equation with electromagnetic source. To be more precise, we consider the plasma in an equilibrium state, and introduce the covariant relativistic version of the Fermi-Dirac distribution function; then, as the first limiting case, we study the Boltzmann plasma; the second limiting case is the degenerated electron gas with the temperature $T_0=0$.

When we consider the potential of the pseudoscalar (axion) field, we obey the following logic. The term (axion), given in parentheses, means that we deal with pseudo-Goldstone bosons, and the appropriate field potential is the specific periodic potential ${\cal V}_{(\rm P)} = {\cal V}_0 \left[1- \cos{\left(\frac{2\pi \phi}{\Phi(\xi)}\right)}\right]$. When we use the term pseudoscalar field, we keep in mind more general idea, that the field $\phi$ plays the role of a pseudoscalar dilaton in analogy with the known scalar dilaton, so that the scalar $\phi F^{*}_{mn}F^{mn}$ that appeared in \cite{Ni77} is considered as the direct analog of the scalar $\psi F_{mn}F^{mn}$ introduced in \cite{dilaton}. For this idea the modified Higgs-type potential, ${\cal V}_{(\rm H)} = \frac12 \gamma \left[\phi^2 {-} \Phi^2(\xi) \right]^2$ seems to be appropriate. In both cases the basic value $\Phi$ is not the constant, it is the function of the modulus $\xi =\sqrt{\xi_k \xi^k}$ of the time-like Killing vector $\xi^k$, which characterizes the static spacetime formed by the axionic dyon. The scalar $\xi$ predetermines the structure of the equilibrium distribution functions of plasma particles entering the formula for effective local equilibrium temperature. We choose these potentials for modeling, since both potentials: periodic and of the Higgs type, possess the following important properties: when $\phi = \Phi(\xi)$ the potentials themselves and their derivatives vanish, ${\cal V}_{|\phi=\Phi}=0$, ${\frac{\partial {\cal V}}{\partial \phi}}_{|\phi=\Phi}= 0$.

We consider the electrodynamic equations, the kinetic equations and the equation for the pseudoscalar (axion) field to be nonlinearly coupled; the gravitational field in the dyon exterior is assumed to be fixed.

The paper is organized as follows. In Section II, we establish the model, represent the distribution functions as solutions to the extended kinetic equations (see Subsection A), derive key equations for the electromagnetic (see Subsection B) and pseudoscalar (axion) (Subsection C) fields, as well as, present two specific sub-models with $\nu=1$ and $\nu=0$ (see Subsection D), which admit the decoupling of the set of master equations. In Section III, we study the reference model, in which the plasma response vanishes, the electric field is predetermined by the axionically deformed magnetic field, and the pseudoscalar field coincides with the basic value $\phi=\Phi(\xi)$. In Section IV we consider in detail the models of axionically active relativistic Boltzmann electron-ion plasma. In Section V we study the axionically active Boltzmann electron-positron plasma. In Section VI we analyze the polarization of a degenerated zero-temperature electron gas. Section VI contains discussion and conclusions.

\section{The model}

The Einstein-Maxwell-Fermi-Dirac-axion model under consideration can be indicated as the two-level hierarchical model, which contains gravitational, pseudoscalar (axion), electromagnetic fields, and a relativistic multi-component plasma. The first (basic) level of this hierarchical stair is occupied by the gravitational field. The spacetime is assumed to be spherically symmetric and static, and is predetermined by the matter, magnetic and electric field inside the object; the gravity field outside of the object is assumed to be known and to be described by the metric
\begin{equation}
ds^2 = c^2 N(r) dt^2 {-} \frac{1}{N(r)}dr^2 {-} r^2 \left(d\theta^2 {+} \sin^2{\theta} d\varphi^2 \right) \,.
\label{eq4}
\end{equation}
The standard metric coefficient  $N(r)$ in the outer region can be modeled using the Reissner-Nordstr\"om solution \cite{Exact}
\begin{equation}
N(r) = 1 - \frac{r_{g}}{r} + \frac{r_{\mu}^2}{r^2}\,,
\label{eq43}
\end{equation}
where $r_g = \frac{2GM}{c^2}$ is the Schwarzschild radius, and $r_{\mu}^2=\frac{G\mu^2}{c^4}$, where $\mu$ is the magnetic charge of the monopole.
When $r_g>2r_{\mu}$, there are two horizons for this metric:
\begin{equation}
N(r_{\pm}) = 0 \ \Rightarrow  r_{\pm} = \frac12 r_g\left[1 \pm \sqrt{1-\frac{4r^2_{\mu}}{r^2_g}} \right]\,.
\label{2eq43}
\end{equation}
We assume that the radius of the object, $r_0$, exceeds the radius of outer horizon $r_{+}$; we consider the plasma and electromagnetic field in the region $r>r_0>r_{+}$.
The spacetime with such metric admits the existence of the time-like Killing vector $\xi^i = {\cal B} \delta^i_0$, where ${\cal B}$ is arbitrary constant. The  modulus of the four-vector $\xi^i$ can be expressed in terms of the metric function $N(r)$ as follows:
\begin{equation}
\xi(r) \equiv \sqrt{g_{mn} \xi^m \xi^n} = {\cal B} \sqrt{N(r)} \,.
\label{eq51}
\end{equation}
Three coupled elements are arranged on the second level of the hierarchical stair: the plasma, pseudoscalar and electromagnetic fields. We assume these elements inherit the symmetry of the surrounding gravitational field; below we formulate mathematically this requirement.

\subsection{Relativistic axionically active plasma in the equilibrium state: Kinetic description}

\subsubsection{Covariant relativistic kinetic equations for the multi-component plasma}

For the description of the state of relativistic multi-component plasma we use the set of covariant kinetic equations \cite{deGroot,Stewart,Ehlers}
\begin{equation}
\frac{p^i}{m_{\rm a}c} \hat{\nabla}_i f_{\rm a} +  \frac{\partial}{\partial p^i} \left({\cal F}^i_{\rm a} f_{\rm a}\right) = \sum_{b} I_{{\rm a}{\rm b}} \,.
\label{kin1}
\end{equation}
Here $f_{\rm a}$ is the 8-dimensional distribution function describing the particles of the sort "${\rm a}$"; the four-vector $p^i$ stands for the particle momentum, this quantity is considered to be random variable in the kinetic approach. The term $m_{\rm a}$ relates to the particle mass; this quantity depends on the sort index ${\rm a}$. The Cartan derivative
\begin{equation}
\hat{\nabla}_i = \nabla_i - \Gamma^j_{ik} p^k \frac{\partial}{\partial p^j}
\label{kin2}
\end{equation}
contains the standard covariant derivative $\nabla_i$, and symmetric Christoffel symbols $\Gamma^j_{ik} = \Gamma^j_{ki}$. The term $I_{{\rm a}{\rm b}}$ describes the integrals of collision between particles of sorts ${\rm a}$ and ${\rm b}$ (see, e.g., \cite{deGroot} for details).

\subsubsection{Remark about the extension of the Lorentz force}

The four-vector ${\cal F}^i_{\rm a}$ in (\ref{kin1}) presents the force, which acts on the particle of the sort ${\rm a}$; we use here the extended Lorentz force, which has the form
\begin{equation}
{\cal F}^i_{\rm a} \equiv \frac{e_{\rm a}}{m_{\rm a} c^2} \left[F^i_{\ k} + \nu \phi F^{*i}_{\ \ k} \right]p^k  \,.
\label{kin3}
\end{equation}
In (\ref{kin3}) $e_{\rm a}$ is the electric charge of particles of the sort ${\rm a}$; $F_{mn}$ is the Maxwell tensor and $F^{* ik} \equiv \frac12 \epsilon^{ikmn} F_{mn}$ is its dual tensor; the Levi-Civita (pseudo)tensor $\epsilon^{ikmn}=\frac{E^{ikmn}}{\sqrt{-g}}$ contains the absolutely skew-symmetric symbol $E^{ikmn}$ and the square root of the determinant $g = \det{g_{ik}}$.
The function $\phi$ describes the dimensionless pseudoscalar (axion) field. The quantity $\nu$ is the dimensionless parameter. When $\nu =0$ we deal with the standard Lorentz force; when $\nu = 1$ the extended Lorentz force (\ref{kin3}) is symmetric with respect to replacement $F_{ik}$ with  $\phi F^{*}_{ik}$; when $0<\nu<1$, this parameter can be absorbed by the pseudoscalar field $\phi$, and we deal, effectively, with the case $\nu \to 1$. For arbitrary $\nu$ the effective force is of the gyroscopic type, i.e., the divergence  $\frac{\partial}{\partial p^i} {\cal F}^i_{\rm a}$ is equal to zero identically. This means that there are two equivalent representations
\begin{equation}
\frac{\partial}{\partial p^i} \left({\cal F}^i_{\rm a} f_{\rm a}\right)  \ = \ {\cal F}^i_{\rm a}  \left(\frac{\partial f_{\rm a}}{\partial p^i} \right)
\label{3kin1}
\end{equation}
of the force terms in the kinetic equations (\ref{kin1}).
There are at least three motives to consider the generalization of the standard Lorentz force.

\vspace{3mm}
\noindent
{\it (i) Intrinsic symmetry of the vacuum Faraday -  Maxwell Electrodynamics in the Minkowski spacetime}.

\noindent
This topic is well documented (see, e.g., \cite{Jackson}); when we formally replace  the Maxwell tensor $F_{ik}$ with the dual tensor $-F^*_{ik}$, the vacuum electrodynamic equations $\nabla_k F^{ik}{=}0$ and $\nabla_k F^{*ik}{=}0$ convert to one another. Moreover, when we search for the potential $A_i$ from the equation $\partial_i A_k{-}\partial_k A_i = F_{ik}$, and thus are faced with six equations for four unknown functions, the Jacobi relationships $\partial_{(l} F_{ik)} {=}0 \rightarrow \nabla_k F^{*ik}=0$ guarantee their compatibility. In turn, when we try to find a dual potential ${\cal A}_i$ using six equations $\partial_i {\cal A}_k{-}\partial_k {\cal A}_i = F^{*}_{ik}$, the corresponding first series of integrability conditions take the form $\partial_{(l} F^{*}_{ik)} {=}0 \rightarrow -\nabla_k F^{ik}{=}0$, thus guaranteeing the existence of the vector field ${\cal A}_i$. This intrinsic symmetry is the first hint for the Lorentz force generalization.

\vspace{3mm}
\noindent
{\it (ii) Hypothesis about the hidden chirality of the electrodynamic vacuum}.

\noindent
When one considers the physical vacuum as a quasi-medium, one deals with the electromagnetic field strength $F_{mn}$ and the field induction $H^{ik}$. In the framework of linear electrodynamics these two quantities are linked by the constitutive equations $H^{ik}=C^{ikmn}F_{mn}$, where $C^{ikmn}$  is the linear response tensor (see, e.g., the book \cite{HehlObukhov} for details). In order to describe the electrodynamic vacuum, one can decompose the quantity $C^{ikmn}$ as
\begin{equation}
C^{ikmn} =  \frac{1}{2\mu_0} \delta^{ik}_{pq}g^{mp} g^{nq} + \frac12 \phi_0 \epsilon^{ikmn}
\label{remark1}
\end{equation}
using two fundamental geometric objects: the Kronecker tensor $\delta^{ik}_{pq}$ and the Levi-Civita tensor $\epsilon^{ikmn}$. In the standard system of units the magnetic permeability of vacuum is equal to one $\mu_0=1$; the magnetoelectric constant $\phi_0$ is unknown, so that $H^{ik}=F^{ik}{+}\phi_0 F^{*ik}$. When $\phi_0$ is constant, the electrodynamic equations
\begin{equation}
\nabla_k H^{ik}{=}0 \,, \nabla_k F^{*ik} {=}0 \rightarrow \nabla_k F^{ik}{=}0 \,, \nabla_k F^{*ik} {=}0
\label{remark2}
\end{equation}
ignore the presence of $\phi_0$, i.e., the vacuum magnetoelectric constant is doomed to be hidden, thus providing {\it potentially} the chirality of the electrodynamic vacuum. When we consider the axionically active vacuum with the pseudoscalar (axion) field $\phi$ instead of the constant $\phi_0$, this hidden chirality of the vacuum becomes explicit \cite{HehlObukhov}. This is the second hint to introduce the generalization of the Lorentz force, and this ansatz allows us to consider the pseudoscalar $e_{\rm a}\phi$ as a pseudo-electric charge, which appears in (\ref{kin3}) in front of the dual Maxwell (pseudo)tensor $F^{*ik}$.

\vspace{3mm}
\noindent
{\it (iii) Hypothesis about the axionically induced magnetic conductivity}.

\noindent
The effect of electric conductivity is well-known; it can be described in terms of electric current $I^i = \sigma E^i$, which is proportional to the electric field four-vector $E^i \equiv F^{ik}U_k$. Here $U^k$ is the macroscopic velocity four-vector attributed to the (quasi)medium, and $\sigma$ is the electric conductivity parameter. When we deal with the axion electrodynamics of conductive medium, electrically neutral as a whole, we can rewrite the first subsystem of the master equations as follows:
$$
\nabla_k H^{ik}{=}-\frac{4\pi}{c} I^i  \rightarrow
$$
\begin{equation}
\nabla_k F^{ik}{=}-\frac{4\pi \sigma}{c} F^{ik}U_k - F^{*ik} \nabla_k \phi \,,
\label{remark3}
\end{equation}
(see. e.g., \cite{Ni77,HehlObukhov,BG,Symm}). Since the gradient four-vector $\nabla_k \phi$ contains the longitudinal part $U_k D\phi \equiv U_k U^l\nabla_l \phi$, and since the four-vector $B^i=F^{*ik}U_k$ describes the magnetic induction in the medium, we can consider the pseudoscalar $\frac{c}{4\pi}D\phi$ as the parameter of axionically induced magnetic conductivity in analogy with the electric conductivity $\sigma$.

Based on these three examples extracted from {\it macroscopic} electrodynamics, we assume that the dynamics of charged particles described on the {\it microscopic} level can be characterized by the extended Lorentz force including the product of dual Maxwell tensor and of pseudoscalar (axion) field (\ref{kin3}). Of course, it is our {\it ansatz}, and the experimental (or observational) data that could clarify only whether the extra parameter $\nu$ is non-vanishing. Finally, we would like to add that in \cite{BP2015} we considered a series of versions of description of the axionically modified Lorentz force; we studied the generalizations of the Bargmann-Michel-Telegdi model, and the non-minimally extended forces. In this work we restrict our-selves by the model with the force (\ref{kin3}).

\subsubsection{Collision integrals and equilibrium state}

We consider below elastic collisions between relativistic fermions only. The corresponding collision integrals are assumed to be in the following standard form:
$$
I_{{\rm a}{\rm b}} = \int \int \int d P' d Q d Q' W(p,q|p',q') \delta(p{+}q{-}p'{-}q') \times
$$
$$
\times
\left[n_{\rm a}(p)n_{\rm b}(q)(1-n_{\rm a}(p'))(1-n_{\rm b}(q')) -
\right.
$$
\begin{equation}
\left.
- n_{\rm a}(p')n_{\rm b}(q')(1-n_{\rm a}(p))(1-n_{\rm b}(q)) \right] \,.
\label{coll1}
\end{equation}
Here the symbol $p$ relates to the momentum four-vector $p^k$ of the particle of the sort ${\rm a}$ before the collision; $p'$ relates to the momentum after collision; the symbol $q$ corresponds to the particle of the sort ${\rm b}$. The particle number functions $n_{\rm a}(p)$ are proportional to the distribution function $f_{\rm a} = h^{-3}\rho_{\rm a}n_{\rm a}$ with the Planck constant $h$ and degeneration factor $\rho_{\rm a}=2s_{\rm a}{+}1$ (below we use $\rho_{\rm a}{=}2$ for fermions with the spin $s_{\rm a}{=}\frac12$). The quantity  $W(p,q|p'q')$ describes the probability of the event that the collision of the pair of fermions with momenta $p$ and $q$ gives the pair with momenta $p'$ and $q'$. As usual, we assume that
\begin{equation}
W(p,q|p',q')=W(q,p|p',q') =  W(p',q'|p,q) \,.
\label{coll2}
\end{equation}
The symbol $dP$ denotes the invariant measure of integration in the momentum space
\begin{equation}
d P \equiv d^4p \sqrt{-g} \ \delta{\left[p_kp^k-m^2_{\rm a}c^2\right]} {\cal H}(p_k V^k) \,,
\label{coll333}
\end{equation}
where the delta function provides the normalization rule for the momentum four-vector, and the Heaviside function  ${\cal H}(p_k V^k)$ with a macroscopic medium velocity $V^k$ provides the only positive part of the particle energy to be included into the integral (see, e.g., \cite{deGroot} for details).

In the equilibrium state, when $I_{{\rm a}{\rm b}} = 0$, we obtain that  $n_{\rm a}$ is the known Fermi-Dirac function
\begin{equation}
n_{\rm a}(x,p) = \frac{1}{e^{\ \cal U}+1} \,,
\label{coll4}
\end{equation}
containing the function
\begin{equation}
 {\cal U} \equiv -{\cal M}_{\rm a}(x) + \xi_k(x) \ p^k \,,
\label{coll3}
\end{equation}
linear in the particle four-momentum. The quantity ${\cal U}$ includes the set of functions ${\cal M}_{\rm a}(x)$ depending on the sort index.
The four-vector $\xi^k(x)$ is unique for all sorts of particles \cite{deGroot}. The function (\ref{coll4}) turns the left-hand side of (\ref{kin1}) into zero, when
$$
\frac12 p^i p^k \left(\nabla_i \xi_k + \nabla_k \xi_i \right) +
$$
\begin{equation}
+ p^k \left[-\frac{\partial}{\partial x^k} {\cal M}_{\rm a}+ \frac{e_{\rm a}}{c} \xi^i \left(F_{ik}+ \nu \phi F^{*}_{ik} \right) \right] = 0\,.
\label{0eq2}
\end{equation}
When all particles are massive, the first term, quadratic in the particle momenta, vanishes if
\begin{equation}
{\cal L}_{\xi} g_{ik} = \nabla_i \xi_k + \nabla_k \xi_i = 0 \,,
\label{eq3}
\end{equation}
where ${\cal L}_{\xi} g_{ik}$ is the Lie derivative of the metric tensor. In other words, $\xi_k$ has to be associated with the Killing vector \cite{Exact}. In order to provide the convergence of all macroscopic moments of the distribution functions we require, as usual, this vector to be time-like, $\xi^k \xi_k >0$. Keeping in mind the physical sense of the parameters in the distribution functions, we have to choose the multiplier ${\cal B}$ in the modulus of the Killing vector as follows (see (\ref{eq51})):
\begin{equation}
\xi^i = \frac{c}{k_{B} T_0} \delta^i_0 \,.
\label{eq5}
\end{equation}
The parameter $k_B$ is the Boltzmann constant, and the parameter $T_0$ has the dimensionality of temperature.
Then we rewrite the second term in ${\cal U}$ (\ref{coll3}) in the form
\begin{equation}
\xi_k p^k \ \to \frac{c\sqrt{m^2_{\rm a}c^2 {+} p^2}}{k_B T(r)} \,,
\label{eq86}
\end{equation}
using the normalization condition $g^{ik}p_i p_k = m^2_{\rm a}c^2$, and two definitions. First, we introduce the local temperature $T(r)=\frac{T_0}{\sqrt{N(r)}}$; in fact, this quantity
plays the role of a "geometric temperature", since it appears as a scalar reciprocal to the modulus of the time-like Killing vector.
Second,  we define the square of spatial momentum
\begin{equation}
p^2 \equiv \frac{1}{N} ({p^{r}})^2 + r^2 ({p^{\theta}})^2 + r^2 \sin^2{\theta} ({p^{\varphi}})^2 \,,
\label{eq140}
\end{equation}
and write the component $p_0$ of the particle four-momentum as
\begin{equation}
p_0 =  N p^0=  \sqrt{N} \sqrt{m^2_{\rm a}c^2 + p^2} \,.
\label{eq14}
\end{equation}
The second necessary condition, which follows from (\ref{0eq2})
\begin{equation}
\frac{\partial {\cal M}_{\rm a}}{\partial x^k}  = \frac{e_{\rm a}}{c} \xi^i \left(F_{ik}+ \nu \phi F^{*}_{ik} \right) \,,
\label{eq6}
\end{equation}
links the scalars ${\cal M}_{\rm a}$ with the Maxwell tensor, its dual and the axion field. For each sort index, the set (\ref{eq6}) contains four equations for determination of one quantity ${\cal M}_{\rm a}$; these four equations are compatible when the commutator of derivatives vanishes
\begin{equation}
\partial_{[j} \partial_{k]} {\cal M}_{\rm a} = 0 \,.
\label{eq7}
\end{equation}
We assume that the axion field inherits the symmetry of the spacetime, and thus $\pounds_{\xi} \phi =0$. In terms of the Lie derivative the equation (\ref{eq7}) with (\ref{eq6}) can be written in the covariant form
$$
\pounds_{\xi}\left[F_{jk} {+} \nu \phi F^{*}_{jk} \right] =
$$
\begin{equation}
= \nu \xi^l \left[\nabla_{l}(F^{*}_{jk}\phi) {+} \nabla_{j}(F^{*}_{kl}\phi) {+} \nabla_{k}(F^{*}_{lj}\phi) \right] \,.
\label{eq8}
\end{equation}
In the absence of the axion field, i.e., when $\phi=0$, we obtain from (\ref{eq8}) the well-known requirement $\pounds_{\xi} F_{jk} = 0$, which means that the Maxwell tensor also inherits the spacetime symmetry. In the general case, (\ref{eq8}) extends the symmetry inheritance condition by involving the pseudoscalar field in this relationship. When we deal with the spherically symmetric static field configuration, the functions ${\cal M}_{\rm a}$ are considered to depend on the radial variable only, thus the compatibility condition (\ref{eq7}) becomes trivial, and four equations (\ref{eq6}) can be reduced to one, if only two quantities are nonvanishing, $F_{0r}\neq 0$, and $F^{*}_{0r}\neq 0$. In other words, the Maxwell tensor $F_{ik}$ has only two nonvanishing components, $F_{0r}$ and $F_{\theta \varphi}$, which describe the radial static electric and magnetic fields of the axionic dyon, respectively. We use the same ansatz as in \cite{Wilczek87}:
\begin{equation}
F_{\theta \varphi} = \mu \sin{\theta} \,, \quad
F_{0r} = - A^{\prime}_{0}(r) \,.
\label{eq9}
\end{equation}
Here and below the prime denotes the derivative with respect to the indicated argument. Keeping in mind that for the metric (\ref{eq4}) we obtain $\sqrt{-g}= r^2 \sin{\theta}$, we can see that for given field configuration the first subset of Maxwell equations $\nabla_k F^{*ik} = 0$ is satisfied identically:
$$
\nabla_k F^{*ik} = \frac{1}{2r^2 \sin{\theta}}\partial_k \left( E^{ikmn} F_{mn} \right) =
$$
\begin{equation}
= \frac{1}{r^2 \sin{\theta}} \left[\delta^i_0 \ \partial_r F_{\theta \varphi} + \delta^i_{\varphi} \ \partial_{\theta} F_{r0}\right] = 0\,.
\label{eq129}
\end{equation}
According to the formula (\ref{eq6}), for this field configuration  there exists only one nontrivial equation for determination of the  quantity ${\cal M}_{\rm a}$, namely
\begin{equation}
{\cal M}^{\prime}_{\rm a}(r) = - \frac{e_{\rm a}}{k_{B}T_0} \left[A^{\prime}_{0}(r)+  \frac{\nu \mu }{r^2} \phi(r)  \right] \,.
\label{eq111}
\end{equation}
The solution to (\ref{eq111}) is
$$
{\cal M}_{\rm a}(r) = \frac{\tilde{{\cal M}}_{\rm a}(r_{*})}{k_{B}T_0} -
$$
\begin{equation}
{-} \frac{e_{\rm a}}{k_{B}T_0} \left[A_{0}(r){-}A_{0}(r^{*}) {+}  \nu \mu \int_{r_{*}}^r \frac{ dz}{z^2}\phi(z) \right],
\label{eq100}
\end{equation}
where the specific value of the radial variable $r_{*}$ appears as a constant of integration (as an example, one can use the radius of the object, $r_{*}{=}r_0$).
Thus, the Fermi-Dirac equilibrium distribution functions (\ref{coll4}) with (\ref{coll3}) can be reconstructed using (\ref{eq100}) and (\ref{eq86}).
The symbol $\tilde{{\cal M}}_{\rm a}(r_{*})$ stands for the chemical potential of the ensemble of particles of the sort ${\rm a}$ on the sphere with radius $r_*$.

Let us consider two cases in more detail: the relativistic Boltzmann plasma and degenerated electron plasma with temperature equal to zero.

\subsubsection{Equilibrium distribution functions for a relativistic Boltzmann plasma}

When we deal with a relativistic Boltzmann plasma the quantity $e^{\cal U}$ in (\ref{coll4}) is much bigger than one, and the distribution function can be presented in the following standard form:
\begin{equation}
f^{(\rm eq)}_{\rm a} = {\cal A}_{\rm a}(r)    \exp{\left\{{-} \frac{c\sqrt{m^2_{\rm a}c^2 {+} p^2}}{k_B T(r)}\right\}} \,,
\label{B2}
\end{equation}
where the functions ${\cal A}_{\rm a}(r)$ are introduced instead of ${\cal M}_{\rm a}(r)$ by the rule
\begin{equation}
{\rm ln}{\cal A}_{\rm a}(r) = {\cal M}_{\rm a}(r) \,,
\label{B1}
\end{equation}
$$
{\cal A}_{\rm a}(r) = {\cal A}_{\rm a}(r^{*}) \exp{\left\{{-} \frac{e_{\rm a}[A_{0}(r){-}A_{0}(r^{*})]}{k_{B}T_0} \right\}} \times
$$
\begin{equation}
\times \exp{\left\{{-} \frac{e_{\rm a} \nu \mu }{k_{B}T_0}\int_{r^{*}}^r \frac{ dz}{z^2}\phi(z) \right\}}   \,.
\label{B3}
\end{equation}
The first moments of the distribution functions
\begin{equation}
{\cal N}^i_{\rm a} = \int dP f^{(\rm eq)}_{\rm a} \ p^i
\label{3eq12}
\end{equation}
are calculated using integration over the invariant volume in the momentum space $dP$ (\ref{coll333}). The macroscopic velocity $V^i$ in the Heaviside function in (\ref{coll333}) is presented as $V^i= \frac{\xi^i}{\xi}=\delta^i_0 N^{-\frac12}$.
Clearly, the integrals (\ref{3eq12}) vanish, if $i=r$, $i=\theta$, $i=\varphi$, because of the symmetry of the equilibrium distribution functions with respect to $p^{r}$, $p^{\theta}$ and $p^{\varphi}$, respectively. Thus, the first moments  ${\cal N}^i_{\rm a}(r)$ have the structure
\begin{equation}
{\cal N}^i_{\rm a}(r)= \delta^i_0 \frac{4\pi {\cal A}_{\rm a}(r)}{\sqrt{N(r)}} \int_0^{\infty}p^2dp \exp{\left\{-\frac{c \sqrt{m^2_{\rm a}c^2 {+} p^2}}{k_{B}T(r)} \right\}} \,.
\label{eq13}
\end{equation}
The last integration with respect to $p$ gives
$$
{\cal N}^i_{\rm a}(r) = \delta^i_0  \frac{ {\cal N}_{\rm a}(r^{*})}{\sqrt{N(r)}}   \left[\frac{\lambda_{\rm a}(r^{*})}{\lambda_{\rm a}(r)}\right]  \left[\frac{K_2(\lambda_{\rm a}(r))}{K_2(\lambda_{\rm a}(r^{*}))}\right] \times
$$
\begin{equation}
\exp{\left\{\frac{e_{\rm a}[A_{0}(r^{*}) {-} A_{0}(r)]}{k_{B}T_0} \right\}} \exp{\left\{- \frac{e_{\rm a} \nu \mu }{k_{B}T_0}\int_{r^{*}}^r \frac{ dz}{z^2}\phi(z) \right\}} \,.
\label{eq15}
\end{equation}
Here $\lambda_{\rm a}(r) \equiv \frac{m_{\rm a}c^2}{k_{B}T(r)}$ is the dimensionless function of radial variable, which gives the ratio between rest energy $m_{\rm a}c^2$ and local thermal energy $k_{B}T(r)$. The term
\begin{equation}
K_{2}(\lambda) \equiv \frac13 \lambda^2 \int_0^{\infty} dz  \ \sinh^4{z} \ e^{-\lambda \cosh{z}}
\label{eq16}
\end{equation}
relates to the McDonald function (see, e.g., \cite{Bateman,deGroot}).
The quantity ${\cal N}_{\rm a}(r^{*})$ describes the number density of particles of the sort ${\rm a}$ on the sphere of radius $r^{*}$; indeed, if we calculate the square of the four-vector (\ref{eq15}) at $r=r_{*}$, we obtain the following value of this scalar $g_{ik}{\cal N}^i_{\rm a}(r_{*}){\cal N}^k_{\rm a}(r_{*}) = {\cal N}^2_{\rm a}(r^{*})$.

\subsubsection{Equilibrium distribution function for a degenerated electron gas}

When $T_0=0$, the Fermi-Dirac distribution function of electrons is known to convert into the Heaviside function
\begin{equation}
f^{(\rm eq)} = 2 h^{-3} {\cal H}[{\cal E}_{\rm F}(r)-c\sqrt{m^2_{\rm e}c^2 + p^2}]\,,
\label{FD1}
\end{equation}
where the threshold Fermi energy is the following function of the radial variable:
$$
{\cal E}_{\rm F}(r) \equiv \frac{1}{\sqrt{N(r)}}\left\{\tilde{{\cal M}}_{\rm e}(r_{*}) {-} \right.
$$
\begin{equation}
\left.
{-}e \left[A_{0}(r){-}A_{0}(r^{*}) {+}  \nu \mu \int_{r^{*}}^r \frac{ dz}{z^2}\phi(z) \right]\right\} \,.
\label{FD2}
\end{equation}
The four-vector ${\cal N}^i_{\rm e}$ is given now by the formula
\begin{equation}
{\cal N}^i_{\rm e}(r) = \delta^i_0  \frac{ 8 \pi}{3\sqrt{N(r)} h^3} {\cal P}^3_{\rm F}(r) \,,
\label{FD17}
\end{equation}
where the Fermi momentum
\begin{equation}
{\cal P}_{\rm F}(r) = \sqrt{\frac{{\cal E}^2_{\rm F}(r)}{c^2} {-} m^2_{\rm e}c^2} = \left[\frac{3h^3}{8\pi}{\cal N}_{\rm e}(r) \right]^{\frac13}
\label{FD18}
\end{equation}
can be expressed either in terms of Fermi energy (\ref{FD2}), or in terms of scalar of local
electron number density ${\cal N}_{\rm e}(r)$. As for the constant $\tilde{{\cal M}}_{\rm e}(r_{*})$, it can be presented as follows
\begin{equation}
\tilde{{\cal M}}_{\rm e}(r_{*}) = c \sqrt{N(r_{*})} \sqrt{m^2_{\rm e}c^2 + \left[\frac{3h^3}{8\pi}{\cal N}_{\rm e}(r_{*}) \right]^{\frac23} } \,,
\label{FD19}
\end{equation}
i.e., it is proportional to the value of the Fermi energy at $r{=}r_{*}$.

\subsection{Master equations of axion electrostatics}

\subsubsection{On the Lagrange formalism}

The basic set of electrodynamic equations can be obtained by the variation procedure with respect to the potential $A_i$ applied to the action functional
$$
S_0 {=} \int d^4 x \sqrt{{-}g} \left\{\frac{R}{2\kappa}
{+} \frac{1}{4} F^{mn}F_{mn}  {+} \frac14 \phi F^*_{mn} F^{mn}
{+} \right.
$$
\begin{equation}
\left. {+}\frac12 \Psi^2_0 \left[- \nabla_k \phi \nabla^k \phi + {\cal V} \right] {+} L_{({\rm
m})} \right\}\,.
\label{axion333}
\end{equation}
Here $R$ is the Ricci scalar; the parameter $\Psi_0$ is connected with the constant of the axion-photon coupling $g_{{\rm A}\gamma \gamma}= \frac{1}{\Psi_0}$ ($g_{{\rm A}\gamma \gamma}<1.47 \cdot 10^{-10} {\rm GeV}^{-1}$, see, e.g., \cite{CAST14}); the quantity ${\cal V}$ is the potential of the pseudoscalar (axion) field.
$L_{({\rm m})}$ is the total Lagrangian of the matter and plasma particles, it can depend, in general case, on the pseudoscalar field $\phi$ and its gradient four-vector $\nabla_p \phi$, on the electromagnetic potential $A_k$ and on the Maxwell tensor $F_{pq}$.
The variation procedure yields
\begin{equation}
\nabla_k \left[F^{ik} {+} \phi F^{*ik}\right] = {-} \frac{4\pi}{c} I^i \,,
\label{eq12}
\end{equation}
where the four-vector of the electric current $I^i$ is defined as follows:
\begin{equation}
I^i \equiv  \frac{c}{4\pi}\left[\frac{\partial L_{({\rm m})}}{\partial A_i} - 2\nabla_p \left(\frac{\partial L_{({\rm m})}}{\partial F_{pi}} \right)  \right] \,.
\label{eq127}
\end{equation}
In the framework of the kinetic approach the structure of the Lagrange function $L_{({\rm m})}$ remains implicit, and the electric current $I^i$ is assumed to be modeled as the sum of the first momenta of the plasma distribution functions:
\begin{equation}
I^i = \sum_{\rm a} e_{\rm a} c \ {\cal N}^i_{\rm a} \,.
\label{3eq1299}
\end{equation}
This four-vector has vanishing divergence, since for the equilibrium plasma
$$
\nabla_i I^i = \sum_{\rm a} e_{\rm a} c \ \int dP p^i \hat{\nabla}_i f^{(\rm eq)}_{\rm a} =
$$
\begin{equation}
- \sum_{\rm a} e_{\rm a} c \ \int dP \frac{\partial}{\partial p^i} \left({\cal F}^i_{\rm a} f_{\rm a}\right) = 0 \,.
\label{3eq129}
\end{equation}
Based on this assumption about the structure of the plasma current, we obtain self-consistently the following nonlinear equations of axion electrodynamics in the axionically active plasma:
\begin{equation}
\nabla_k F^{ik} = -  F^{*ik}\nabla_k \phi - 4\pi \sum_{\rm a} e_{\rm a} \int dP f^{(\rm eq)}_{\rm a} \ p^i \,.
\label{6eq12}
\end{equation}
The first momenta of the distribution functions are calculated above; below we draw attention to the key equations of plasma electrostatics for two special cases: the Boltzmann plasma and degenerated electron gas.

\subsubsection{Key electrostatic equation for the Boltzmann plasma}

For the Boltzmann plasma with the distribution function (\ref{B2}) and its first moment (\ref{eq15}), the key equation of the axion electrostatics takes the form:
$$
\frac{1}{r^2}\frac{d}{dr} \left\{r^2 \frac{d}{dr}\left[A_0(r)-A_0(r^{*}) \right] + \mu \phi \right\}
=
$$
$$
=- \frac{4\pi}{\sqrt{N(r)}} \sum_{\rm a} e_{\rm a} {\cal N}_{\rm a}(r^{*}) \left[\frac{\lambda_{\rm a}(r^{*})}{\lambda_{\rm a}(r)}\right] \ \left[\frac{K_2(\lambda_{\rm a}(r))}{K_2(\lambda_{\rm a}(r^{*}))}\right] \times
$$
\begin{equation}
\exp{\left\{\frac{e_{\rm a}[A_{0}(r^{*}){-}A_{0}(r)]}{k_{B}T_0} \right\}} \exp{\left\{{-} \frac{e_{\rm a} \nu \mu }{k_{B}T_0}\int_{r^{*}}^r \frac{ dz \phi(z)}{z^2} \right\}} \,.
\label{eq17}
\end{equation}
Clearly, it is convenient to make the following definition
\begin{equation}
\Theta(r) \equiv A_{0}(r) - A_{0}(r^{*}) \,, \quad \Theta(r_{*}) = 0 \,.
\label{def1}
\end{equation}
Also, we assume the electric field to have the Coulomb-type asymptote, i.e.,
\begin{equation}
\frac{d A_{0}(r)}{dr} (r \to \infty)  \propto \frac{1}{r^{2+\varepsilon}} \to 0 \,,
\label{def11}
\end{equation}
where the effective parameter $\varepsilon$ is non-negative, $\varepsilon \geq 0$.

\subsubsection{Key electrostatic equation for degenerated electron gas}

When we deal with the distribution function (\ref{FD1}) and its first moment (\ref{FD17}), we obtain the following key electrostatic equation:
$$
\frac{1}{r^2}\frac{d}{dr} \left[r^2 \frac{d \Theta}{dr} + \mu \phi \right]
=
$$
$$
={-} \frac{32 e \pi^2}{c^3 h^3 N^2(r)} \left\{\left[c\sqrt{N(r_{*})} \sqrt{m^2_{\rm e}c^2 + {\cal P}^2_{\rm F}(r_{*}) } {-}
\right. \right.
$$
\begin{equation}
\left. \left.
- e\Theta(r) {-} e \nu \mu \int_{r_{*}}^r \frac{dz}{z^2} \phi(z)\right]^2 {-}
m^2_{\rm e}c^4 N(r) \right\}^{\frac32}
 \,.
\label{FD27}
\end{equation}
One specific detail appears in this equation in comparison with the one obtained for the Boltzmann plasma: there exists, generally speaking, at least one sphere of a finite radius, say, $r_{\rm F}$, for which the right-hand side of the equation (\ref{FD27}) vanishes.
$$
e\left[\Theta(r_{\rm F}) + \nu \mu \int_{r_{*}}^{r_{\rm F}} \frac{dz}{z^2} \phi(z) \right] =
$$
\begin{equation}
=  c\sqrt{N(r_{*})} \sqrt{m^2_{\rm e}c^2 + {\cal P}^2_{\rm F}(r_{*}) }
\pm m_{\rm e}c^2 \sqrt{N(r_{\rm F})}
 \,.
\label{FD71}
\end{equation}
For $r>r_{\rm F}$ the right-hand side of this equation becomes imaginary. In other words, we deal with a layer $r_{*}<r<r_{\rm F}$ in which the degenerated electron gas is arranged; clearly, the distribution of this gas is inhomogeneous, the gas is polarized, and one can say that the compensating electric field is created with participation of the axionically active plasma in the gravitational field.
The analysis of obtained equations of plasma electrostatics is done in the next sections.

\subsection{Master equation for the pseudoscalar (axion) field}

\subsubsection{Consequences of the variation formalism}

Variation of the action functional (\ref{axion333}) with respect to $\phi$ gives the key equation for the axion field in the following form:
\begin{equation}
\nabla^k  \nabla_k \phi + \phi \frac{\partial {\cal V}}{\partial \phi^2}
= - \frac{1}{4 \Psi^2_0} F^{*}_{mn}F^{mn} + {\cal J} \,,
 \label{axion33}
\end{equation}
The pseudoscalar source-type function ${\cal J}$ can be formally obtained as
\begin{equation}
{\cal J} = - \frac{1}{\Psi^2_0}\left[\frac{\partial L_{(\rm m)}}{\partial \phi} - \nabla_p \left(\frac{\partial L_{(\rm m)}}{\partial \nabla_p \phi} \right) \right] \,,
 \label{axion33q}
\end{equation}
and should be modeled as a pseudo-moment of the distribution function of the zero order. The kinetic approach for the relativistic plasma description is equipped by intrinsic moments of an arbitrary order:
\begin{equation}
T^{i_1...i_s}_{\rm a} =  \int dP f^{(\rm eq)}_{\rm a} p^{i_1} \cdot \cdot \cdot p^{i_s} \,.
 \label{axion33qw}
\end{equation}
The moment of the second order $T^{ik}_{\rm a}$ stands for description of the stress-energy tensor of the plasma particles of the sort ${\rm a}$; the moment of the first order relates to the particle number four vectors; zero-order moment is usually associated with the trace of the stress - energy tensor. In this multi-moment scheme there is no intrinsic pseudo-moment. Based on this fact, we consider a special ansatz for the structure of the term $L_{(\rm m)}$ in (\ref{axion333}), namely, we assume that the pseudoscalar source-like term ${\cal J}$ introduced by (\ref{axion33q}) is vanishing.

\subsubsection{Key equation for the pseudoscalar (axion) field}

With the ansatz for the term ${\cal J}$ the equation (\ref{axion33}) can be rewritten as follows:
\begin{equation}
\frac{d}{dr} \left[r^2 N \frac{d \phi}{dr} + \frac{\mu}{\Psi^2_0} A_0 \right] = \frac12 r^2 \frac{\partial {\cal V}}{\partial \phi} \,.
 \label{axion133}
\end{equation}
As we mention in the Introduction, we deal below with either the periodic axion potential ${\cal V}_{(\rm P)} = {\cal V}_0 \left[1- \cos{\left(\frac{2\pi \phi}{\Phi(\xi)}\right)}\right]$, or with the Higgs potential ${\cal V}_{(\rm H)} = \frac12 \gamma \left[\phi^2 {-} \Phi^2(\xi) \right]^2$ for the pseudoscalar field of general type. In both cases the potentials themselves and their derivatives vanish, when $\phi = \Phi(\xi)$, i.e., ${\cal V}_{|\phi=\Phi}=0$, ${\frac{\partial {\cal V}}{\partial \phi}}_{|\phi=\Phi}= 0$.
In both cases the quantity $\Phi(\xi)$ is not constant, it depends on the metric coefficient $N(r)$, i.e., $\Phi(N(r))$, via the modulus of the Killing vector $\xi$. Equivalently, one can say, that $\Phi$ depends on the equilibrium temperature of the plasma. The values $\phi = \pm \Phi(\xi)$ indicate the local minima of both potentials; the positions and depths of these minima depend on the distance to the center. The difference between minima in these two potentials is that the Higgs potential has only two minima, while the periodic potential has an infinite number of minima at $\phi_{(k)} = \pm k \Phi(\xi)$, where $k$ is the integer.

One can see that the term $\frac{\partial {\cal V}}{\partial \phi}$ in the right-hand side of (\ref{axion133}) plays a similar role for the pseudoscalar (axion) field as the collision integrals $I_{\rm ab}$ play for the plasma (see (\ref{kin1})). In the kinetic approach we consider the equilibrium plasma assuming that $I_{\rm ab}=0$ and the equilibrium functions convert the left-hand sides of the kinetic equations (\ref{kin1}) to zero. When we deal with the equation (\ref{axion133}), we consider the solutions, for which  $\phi(r) = + \Phi(r)$, the right-hand side of this equation vanishes, so the function $\Phi(r)$ satisfies the equation
\begin{equation}
\frac{d}{dr} \left[r^2 N \frac{d \phi}{dr} + \frac{\mu}{\Psi^2_0} A_0 \right] = 0 \,.
 \label{axionvv33}
\end{equation}
The obtained  equation can be integrated immediately, yielding
\begin{equation}
N r^2 \frac{d \Phi}{dr}  + \frac{\mu}{\Psi^2_0} \Theta = K \,, \quad K = \left[N r^2 \frac{d \Phi}{dr}  \right]_{|r=r^{*}} \,.
\label{axion921}
\end{equation}
The constant of integration $K$ is connected with the value of the derivative of the axion field at the reference sphere $r=r_{*}$.

\subsection{Two special versions of the set of key equations for the Boltzmann plasma}

\subsubsection{The case $\nu = 0$: There is no axionic modification of the Lorentz force}

When the Lorentz force has the standard form, i.e., $\nu = 0$, the equation for the potential function $\Theta$
$$
\frac{1}{r^2}\frac{d}{dr} \left(r^2 \frac{d \Theta}{dr} \right)  - \frac{\mu^2}{r^4 N(r)\Psi^2_0} \Theta = - \frac{\mu}{r^4 N(r)} K -
$$
\begin{equation}
- \sum_{\rm a} \frac{4\pi e_{\rm a} {\cal N}_{\rm a}(r^{*})}{\sqrt{N(r)}} \left[\frac{\lambda_{\rm a}(r^{*})K_2(\lambda_{\rm a}(r))}{\lambda_{\rm a}(r) K_2(\lambda_{\rm a}(r^{*}))}\right] \exp{\left[-\frac{e_{\rm a} \Theta}{k_{B}T_0} \right]}
\label{Ieq17}
\end{equation}
happens to be decoupled from the equation for the axionic field.
There are two sources on the right-hand side of this equation: first, the source proportional to the axionic "charge" $K$;  second,  the source provided by plasma.
When the electric potential $\Theta(r)$ is found, the distribution of the pseudoscalar field can be described by the solution to the equation
\begin{equation}
\frac{d \Phi}{dr}   = \frac{1}{r^2 N(r)}\left[K - \frac{\mu}{\Psi^2_0} \Theta(r) \right]\,.
\label{If1}
\end{equation}
The solution to (\ref{If1}) can be written in quadratures
$$
\Phi(r)   = \Phi(r_*) + \frac{K}{(r_{+}{-}r_{-})} \ {\rm ln} \left|\frac{(r{-}r_{+})(r_*{-}r_{-})}{(r{-}r_{-})(r_*{-}r_{+})} \right| -
$$
\begin{equation}
-\frac{\mu}{\Psi^2_0} \int_{r_*}^{r} \frac{dz\Theta(z)}{(z-r_{+})(z-r_{-})}\,.
\label{If2}
\end{equation}
(See (\ref{2eq43}) for definition of $r_{+}$ and $r_{-}$). If the electromagnetic field has the Coulombian asymptote, we obtain
\begin{equation}
\Phi(\infty) - \Phi(r_*) = \frac{K}{(r_{+}-r_{-})} \ {\rm ln} \left|\frac{(r_*-r_{-})}{(r_*-r_{+})} \right| \,.
\label{If3}
\end{equation}
This difference is positive when $K>0$.

\subsubsection{The case $\nu = 1$: There is intrinsic symmetry in the axionic modification of the Lorentz force}

When the tensors $F_{ik}$ and $\phi F^{*}_{ik}$ enter the Lorentz force symmetrically, i.e.,  $\nu = 1$, we can, clearly, introduce the super-potential $\Psi(r)$ as follows:
\begin{equation}
\Psi(r) \equiv \Theta(r) +  \mu \int_{r^{*}}^r \frac{ dz \Phi(z)}{z^2}   \,,
\label{IIaxion92}
\end{equation}
so that the corresponding master equation does not contain the axion field explicitly
$$
\frac{1}{r^2}\frac{d}{dr} \left(r^2 \frac{d \Psi}{dr} \right)
=
$$
\begin{equation}
=
-  \sum_{\rm a} \frac{4\pi e_{\rm a} {\cal N}_{\rm a}(r^{*}) \lambda_{\rm a}(r^{*})K_2(\lambda_{\rm a}(r))}{\sqrt{N(r)}\lambda_{\rm a}(r)K_2(\lambda_{\rm a}(r^{*}))}  \exp{\left[-\frac{e_{\rm a}\Psi(r)}{k_{B}T_0} \right]}  \,.
\label{IIeq17}
\end{equation}
The super-potential is assumed to satisfy two conditions:
\begin{equation}
\Psi(r_*)=0 \,, \quad \left(\frac{d \Psi}{dr} \right)_{|r \to \infty} = \lim_{r \to \infty}\left[\frac{\mu  \Phi(r)}{r^2} \right] = 0 \,.
\label{4IIaxion92}
\end{equation}
When the solution for $\Psi(r)$ is found, we can obtain the pseudoscalar field from the equation
\begin{equation}
\frac{1}{r^2} \frac{d}{dr} \left[N r^2 \frac{d \Phi}{dr}\right] - \frac{\mu^2 \Phi}{r^4 \Psi^2_0}  =  - \frac{\mu}{r^2\Psi^2_0} \frac{d \Psi}{dr} \,,
\label{3IIaxion92}
\end{equation}
and then reconstruct $\Theta(r)$ using the relation (\ref{IIaxion92}).

\section{Reference model: The Boltzmann plasma response is absent}

Let us consider, first, the model in which the right-hand side of the equation (\ref{eq17})
is equal to zero. As we show below, this situation can occur, in principle, in at least three cases, first, when one deals with the electron-positron plasma; second, when the generalized Lorentz force contains $F_{mn}$ and $\phi F^*_{mn}$ in a symmetric form; third, when plasma effects are negligible.

\subsection{Key equation for the axion field, and its fundamental solutions}

We deal now with the set of two coupled differential equations
of second order:
\begin{equation}
\frac{d}{dr} \left[r^2 \frac{d \Theta}{dr} {+} \mu \Phi \right] =0 \,,
  \label{001}
\end{equation}
\begin{equation}
\frac{d}{dr} \left[r^2 N \frac{d \Phi}{dr} + \frac{\mu}{\Psi^2_0} \Theta \right] = 0  \,,
 \label{0001}
\end{equation}
and, clearly, the equation for the axion field can be decoupled. Indeed, the electric potential can be expressed in terms of derivative of the axion field using (\ref{0001}):
\begin{equation}
\Theta(r) = \frac{\Psi^2_0}{\mu}\left[r^2_{*}N(r_{*}) {\frac{d \tilde{\Phi}}{dr}}_{|r=r_{*}} - r^2 N(r) \frac{d \tilde{\Phi}}{dr} \right] \,.  \label{003}
\end{equation}
This formula contains the quantity $\tilde{\Phi}$ defined as
\begin{equation}
\tilde{\Phi} = \Phi - \frac{{\cal M}}{\mu} \,, \quad {\cal M} \equiv \lim_{r \to \infty}\left[\mu \Phi(r) + r^2 \frac{d\Theta}{dr} \right] \,.  \label{0032}
\end{equation}
We assume that this quantity is finite.
In these terms the key equation for the axion field takes the form
\begin{equation}
r^2 \frac{d}{dr} \left(r^2 N \frac{d \tilde{\Phi}}{dr}\right) = \frac{\mu^2}{\Psi^2_0} \tilde{\Phi} \,,
 \label{004}
\end{equation}
its solution gives the profile of the  axion field in the outer zone of the magnetic monopole.
The key equation (\ref{004}) can be analyzed as follows.
First of all, we introduce the dimensionless radial variable $x$, using the relationships
\begin{equation}
x = \frac{r}{r_{+}} \,, \quad r_{+} = \frac12 r_g\left[1+ \sqrt{1-\frac{4r^2_{\mu}}{r^2_g}} \right] \,,
\label{au1}
\end{equation}
and add two convenient  parameters
\begin{equation}
a \equiv \frac{r_g}{r_{+}} -1 = \frac{r_{-}}{r_{+}} < 1 \,, \quad r_{\rm A} \equiv \frac{\mu}{\Psi_0} \,.
\label{au11}
\end{equation}
There is, of course, the limiting (extremal) case, where
\begin{equation}
2r_{\mu}= r_g  \ \ \rightarrow    r_{+} = r_{-} \equiv r_{h} = r_g \,, \quad a =1 \,,
\label{qqq}
\end{equation}
but below we consider only the case $a<1$. The parameter $r_{\rm A}$ describes a new distance, which characterizes the interaction between the magnetic field of the monopole and axionic field.

Since in our model $r>r_{+}$, we see that $x>1$.
In these terms, the equation (\ref{004}) can be rewritten as
\begin{equation}
\frac{d^2 \tilde{\Phi}}{dx^2} {+} \frac{d \tilde{\Phi}}{dx} \left[\frac{1}{x{-}1}{+} \frac{1}{x{-}a} \right] {-}  \left(\frac{r^2_{\rm A}}{r^2_{+}}\right) \frac{\tilde{\Phi}}{x^2(x{-}1)(x{-}a)} = 0 \,.
\label{Fu29}
\end{equation}
As usual, the solution to the equation (\ref{Fu29})
\begin{equation}
\tilde{\Phi}(x) = C_1 Y_{(1)}(x) + C_2 Y_{(2)}(x)
\label{Fu31}
\end{equation}
is a linear combination of two fundamental solutions $Y_{(1)}(x)$ and $Y_{(2)}(x)$.
The Wronsky determinant for these fundamental solutions is of the form
\begin{equation}
W\left[Y_{(1)}, Y_{(2)}\right] \equiv W(x) = W(x_{*}) \frac{(x_* {-}1)(x_* {-}a)}{(x-1)(x-a)}  \,,
\label{Fu99}
\end{equation}
where
\begin{equation}
W(x_*) = Y_{(1)}(x_*) Y^{\prime}_{(2)}(x_*)-Y^{\prime}_{(1)}(x_*)Y_{(2)}(x_*) \,.
\label{Fu999}
\end{equation}
For the determination of the integration constants, we can use the formulas
$$
C_1 = \frac{\tilde{\Phi}(x_{*})Y^{\prime}_{(2)}(x_*)- \tilde{\Phi}^{\prime}(x_*)Y_{(2)}(x_*)}{W(x_*)}
 \,,
$$
\begin{equation}
 C_2 = \frac{\tilde{\Phi}^{\prime}(x_*)Y_{(1)}(x_*) -\tilde{\Phi}(x_{*})Y^{\prime}_{(1)}(x_*) }{W(x_*)}
\,.
\label{Fu37}
\end{equation}
Here and below we use the convenient quantity $x_{*} \equiv \frac{r_{*}}{r_{+}}$.

\subsection{Remark concerning the Fuchs-type functions}

The equation (\ref{Fu29}) can be obtained from the Fuchs-type equation
$$
Y^{\prime \prime}(x) + Y^{\prime} (x) \left[\frac{\gamma}{x} +  \frac{\delta}{x-1}+\frac{\epsilon_1}{x-a}+\frac{\epsilon_2}{x-b}\right] +
$$
\begin{equation}
+ Y \frac{\left(\alpha \beta x^2 + p_1 x + p_2\right)}{x(x-1)(x-a)(x-b)} = 0 \,,
\label{Fu1}
\end{equation}
(see, e.g., \cite{Bateman,Ince,Poole}), if to choose the parameters as follows:
\begin{equation}
b{=} \alpha{=}p_1{=}0 \,, \quad  \epsilon_2{=} {-}\gamma \,,  \quad \delta {=} \epsilon_1 {=} 1 \,, \quad p_2 = -\frac{r^2_{\rm A}}{r^2_{+}} \,.
\label{Fu2}
\end{equation}
Using the Fuchsian relation (see, e.g., \cite{Ince,Poole} for motivation of this requirement)
\begin{equation}
\alpha + \beta + 1 = \gamma + \delta + \epsilon_1 + \epsilon_2  \,,
\label{Fu3}
\end{equation}
we obtain additionally $\beta=1$; the parameter $\gamma$ remains hidden. This means that in the model under consideration the pseudoscalar field $\tilde{\Phi}$ is described by the Fuchs-type function with the parameters (\ref{Fu2}). Also this function can be indicated as generalized Heun's function (see, e.g., \cite{Heun} and references therein); the corresponding equations are well studied in the context of searching for generalized spherical wave functions.

Typical sketches for the fundamental solutions $Y_{(1)}(x)$ and $Y_{(2)}(x)$ are presented in Figs.1 and 2.

\begin{figure}
	\includegraphics[width=90mm,height=70mm]{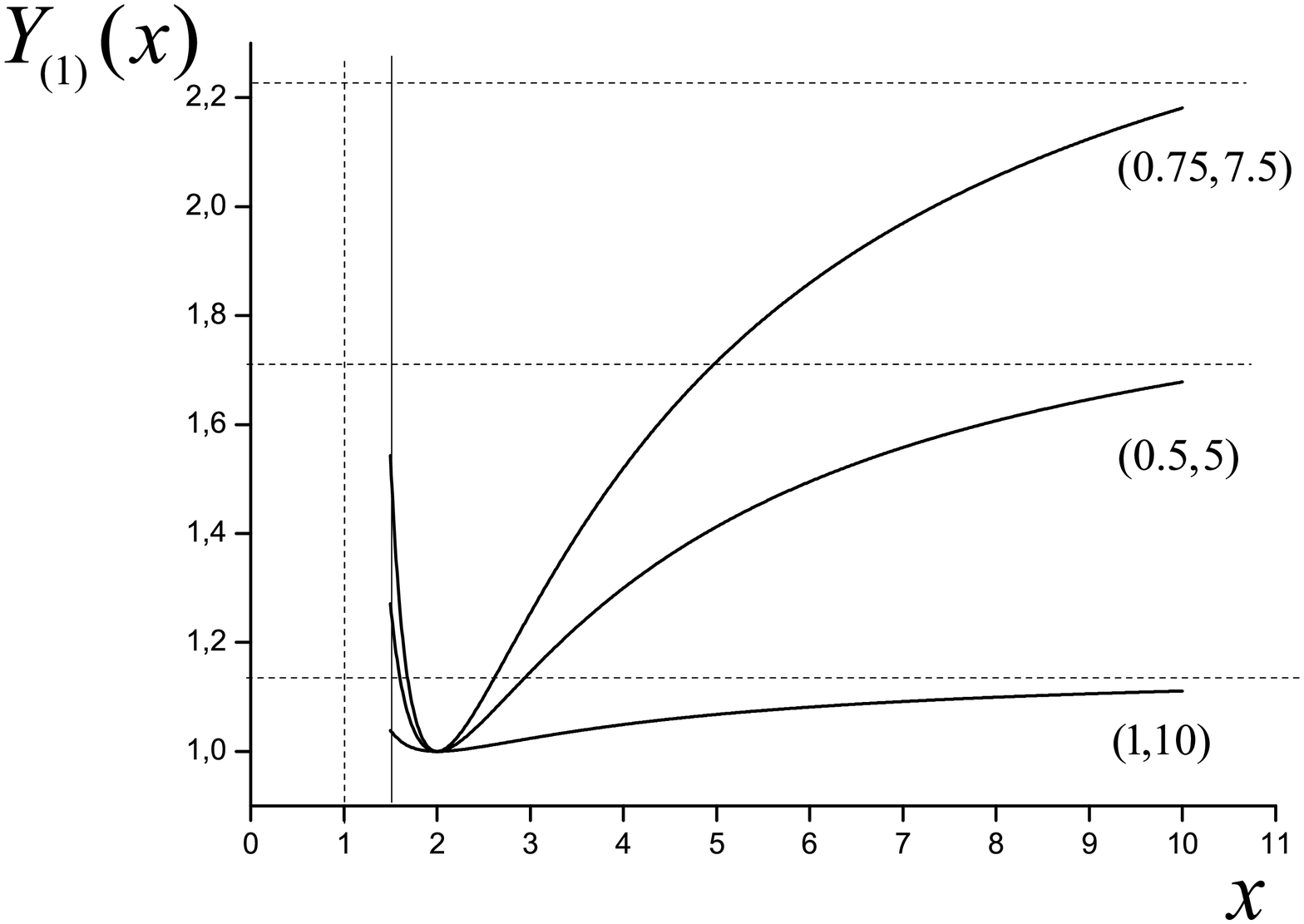}
	\caption{Plot of the Fuchs-type function $Y_{(1)}(x,a,p_2)$, the first fundamental solution to the equation (\ref{Fu29}), which corresponds to the standard boundary conditions $Y_{(1)}(x_*){=}1$, $Y^{\prime}_{(1)}(x_*){=}0$.
For illustration, we depict three curves corresponding to three sets of the parameters $(a,p_2)$ (these quantities are indicated on the plot near the corresponding curves). We put $x_{*}{=}2$. The vertical line $x{=}1$ relates to the outer horizon; the value $x{=}x_0{=}1.5$ relates to the radius of the dyon. The function is monotonic for $x>x_{*}{=}2$. The profiles of the function $Y_{(1)}(x,a,p_2)$ have horizontal asymptotes at $x \to \infty$; these asymptotic values are indicated on the plot by the corresponding horizontal lines.
}
\end{figure}

\begin{figure}
	\includegraphics[width=80mm,height=70mm]{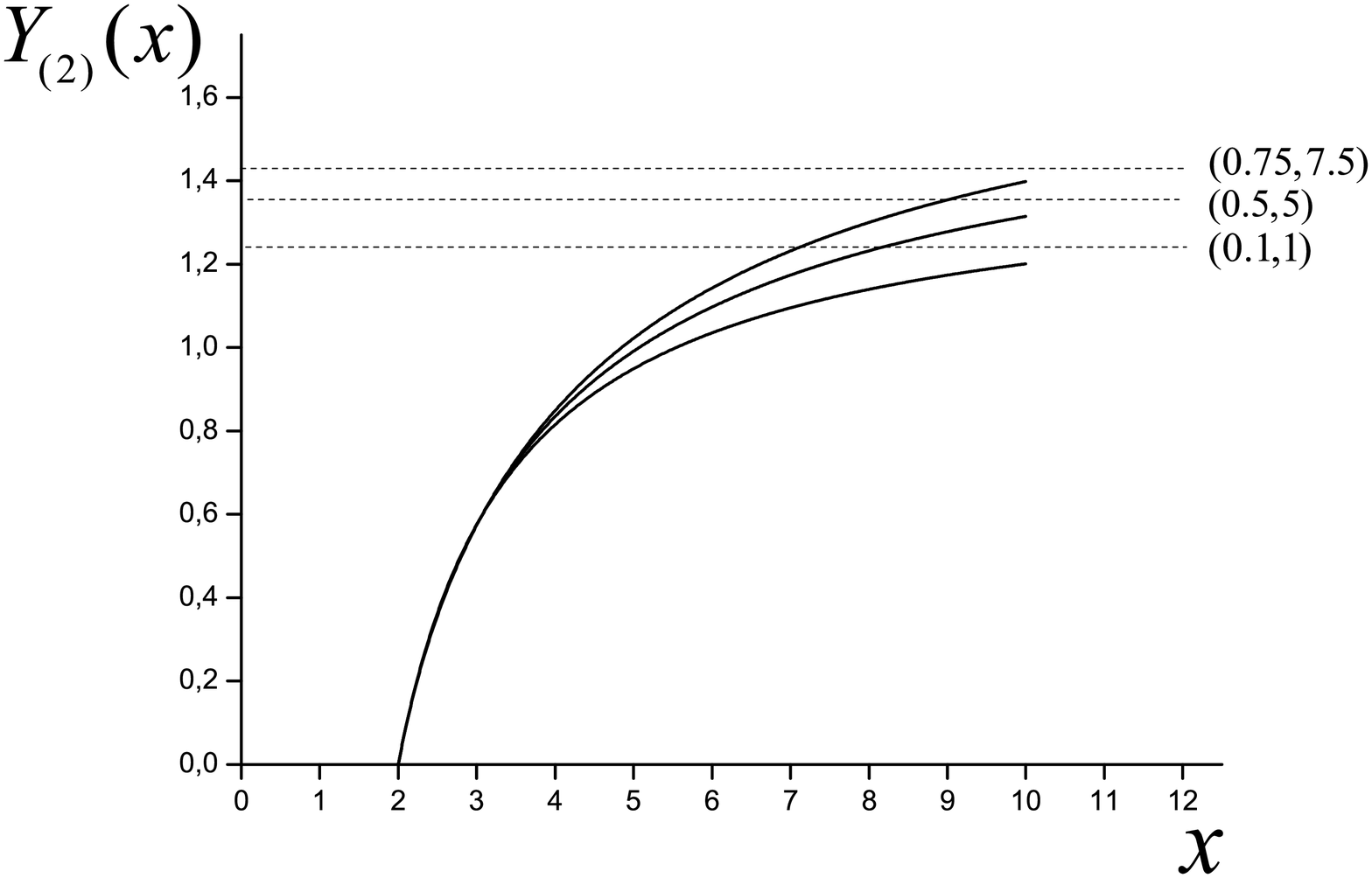}
\caption{Plot of the Fuchs-type function $Y_{(2)}(x,a,p_2)$, the second fundamental solution to the equation (\ref{Fu29}), which corresponds to the standard boundary conditions $Y_{(2)}(x_*){=}0$, $Y^{\prime}_{(1)}(x_*){=}1$.
For illustration, we depict three curves corresponding to three sets of the parameters $(a,p_2)$ in the outer domain $x \geq x_{*}{=}2>x_0 $. The function is monotonic; the profiles of the function $Y_{(2)}(x,a,p_2)$ have horizontal asymptotes at $x \to \infty$, indicated by the corresponding horizontal lines.}
\end{figure}

\section{Relativistic Boltzmann electron - ion plasma}

In this Section we assume that plasma contains electrically charged particles with different masses $m_{\rm a}$, our purpose is to describe the total plasma polarization produced both: by coupling photons to axions, and by coupling of massive fermions to the gravity field.

\subsection{Plasma with the standard Lorentz force, $\nu=0$}

In the regime, when the electrostatic energy per one particle is much smaller than its average kinetic energy, i.e., $\left|\frac{e_{\rm a}\Theta}{k_{B}T_0}\right|<<1$, one can represent the exponential term in (\ref{Ieq17}) by two first elements of decomposition, so the master equation (\ref{Ieq17}) can be transformed into
\begin{equation}
\frac{1}{r^2}\frac{d}{dr} \left(r^2 \frac{d \Theta}{dr} \right)  {-} \left[\frac{1}{{\cal R}^2_{*}} {+} \frac{\mu^2}{r^4 N(r)\Psi^2_0}\right] \Theta =
{\cal I}_{*} {-} \frac{\mu  K}{r^4 N(r)}\,.
\label{plus1}
\end{equation}
In the right-hand side of this equation there are two electric sources. The first one
\begin{equation}
{\cal I}_{*} \equiv -  \sum_{\rm a}\frac{4\pi  e_{\rm a} {\cal N}_{\rm a}(r^{*})}{\sqrt{N(r)}} \left[\frac{\lambda_{\rm a}(r^{*})K_2(\lambda_{\rm a}(r))}{\lambda_{\rm a}(r)K_2(\lambda_{\rm a}(r^{*}))}\right] \,,
\label{axion6}
\end{equation}
is produced by the plasma polarization in the gravitational field and is, in fact, the local density of the electric charge in the axionically active polarized plasma.  The second source is proportional to the effective axionic charge $K$. Also, we introduce the following auxiliary function:
\begin{equation}
\frac{1}{{\cal R}^2_{*}} \equiv  \sum_{\rm a} \frac{4\pi e^2_{\rm a} {\cal N}_{\rm a}(r^{*})}{k_{B}T_0 \sqrt{N(r)}} \left[\frac{\lambda_{\rm a}(r^{*})K_2(\lambda_{\rm a}(r))}{\lambda_{\rm a}(r)K_2(\lambda_{\rm a}(r^{*}))}\right]  \,,
\label{axion5}
\end{equation}
where the function ${\cal R}_{*}(r)$ describes the local (depending on the position) screening radius.

\subsection{Recovering the standard Pannekoek-Rosseland effect}

When the axion field is absent, and $\mu=0$, there is only one source ${\cal I}_{*}$ in the right-hand side of the equation  (\ref{plus1}); first of all, we consider the non-relativistic version of this source in order to recover the known results.
Non-relativistic limit means that $\lambda_{\rm a}>>1$, and we can use for calculations the leading-order term in the asymptotic decomposition of the McDonald function:
\begin{equation}
K_{2}(\lambda)  \to \sqrt{\frac{\pi}{2\lambda}} e^{-\lambda} \left[1 + \frac{15}{8\lambda}  + ... \right] \,.
\label{eq163}
\end{equation}
When we consider the plasma far from horizon, i.e., when  $r > r_{*} >> r_{+}$,
one can obtain the classical formula for the local charge density in plasma:
\begin{equation}
{\cal I}_{*} \simeq {-}  4\pi \sum_{\rm a}  e_{\rm a} {\cal N}_{\rm a}(r^{*}) \exp\left[\frac{GM m_{\rm a}}{k_B T_0} \left(\frac{1}{r} {-} \frac{1}{r_{*}} \right)\right] .
\label{axion93}
\end{equation}
In this limit the quantity $\frac{1}{{\cal R}^2_{*}}$ has the form
\begin{equation}
\frac{1}{{\cal R}^2_{*}} \to  \sum_{\rm a} \frac{4\pi e^2_{\rm a} {\cal N}_{\rm a}(r^{*})}{k_{B}T_0} \equiv \frac{1}{\lambda^2_{\rm D}} \,,
\label{axion92}
\end{equation}
where $\lambda_{\rm D}$ is the classical Debye radius \cite{deGroot}.
In particular, when one deals with the layer near Earth's surface, $r=r_{*} {=} R_0$, and $z$ is the altitude $z \simeq r-R_0<<R_0$, one can consider the model equation
for the electric potential written as follows:
\begin{equation}
\frac{d^2 \Theta}{dz^2} - \frac{\Theta}{\lambda^2_{\rm D}} = -  4\pi \sum_{\rm a}  e_{\rm a} {\cal N}_{\rm a}(R_0) \exp{\left(- \frac{m_{\rm a} g z}{k_B T_0} \right)} \,,
\label{axion94}
\end{equation}
where $g {=} \frac{GM}{R^2_0}$ is the acceleration of free fall on Earth's surface, $A_0(R_0)=0$.
The local electro-neutrality is violated and the electric potential $\Theta(z)$, which satisfies the conditions $\Theta(0)=0$, $\Theta(\infty)=0$, is of the following form
\begin{equation}
\Theta(z) = 4\pi \sum_{\rm a} \frac{e_{\rm a} {\cal N}_{\rm a}(R_0) \lambda^2_{\rm D}}{\left(1 {-} \lambda^2_{\rm D} \alpha^2_{\rm a}\right)}
\left[e^{-\alpha_{\rm a} z} {-} e^{{-}\frac{z}{\lambda_{\rm D}}}\right] \,,
\label{axion95}
\end{equation}
where $\alpha_{\rm a} \equiv \frac{m_{\rm a} g}{k_B T_0}$. Various physical consequences of this formula are discussed in literature (see, e.g., \cite{P,R,1,2,3,4,5,6}).

\subsection{High-temperature plasma}

When we consider the plasma configurations in astrophysical systems with strong gravitational and magnetic fields, we deal, in fact, with particles whose kinetic energy (on average) is of the order or exceeds the rest energy, $k_{B}T_0 > m_{\rm a} c^2$. For numerical calculations we use below the whole relativistic representation of the electric source (\ref{axion6}), however, in order to have an analytical progress in our calculations, in this Section we assume that $\lambda_{\rm a} = \frac{m_{\rm a}c^2}{k_{B}T_{0}} <<1$.
Asymptotic behavior of the function $K_2(\lambda_{\rm a})$ at $\lambda_{\rm a} \to 0$ is characterized by the well-known principal limit (see, e.g., \cite{deGroot})
\begin{equation}
\lim_{\lambda_{\rm a} \to 0} \left[\lambda^2_{\rm a}K_{2}(\lambda_{\rm a})\right]  =  2 \,.
\label{MD1}
\end{equation}
For our purposes we need at least two terms of the asymptotic decomposition; the extended procedure of this decomposition is the following. First,
we consider the definition of the McDonald function with integer index $\rho=2$ as the limit \cite{Bateman}
$$
K_2(z) \equiv \lim_{\rho \to 2} \left\{\frac{\pi}{2 \sin{\pi \rho}} \sum_{k=0}^{\infty} \frac{1}{k!} \left(\frac{z}{2}\right)^{2k}  \times \right.
$$
\begin{equation}
\left.
\times \left[  \frac{\left(\frac{z}{2}\right)^{-\rho}}{\Gamma(k{-}\rho{+}1)} -
 \frac{\left(\frac{z}{2}\right)^{\rho}}{\Gamma(k{+}\rho{+}1)}
\right] \right\} \,,
\label{MD2}
\end{equation}
and transform this limit by the l'Hospital rule
$$
K_2(z)=\lim_{\rho \to 2} \left\{\frac{1}{2 \cos{\pi \rho}} \sum_{k=0}^{\infty} \frac{1}{k!} \left(\frac{z}{2}\right)^{2k}  \times
\right.
$$
\begin{equation}
\left.
\times \left[ \left(\frac{z}{2}\right)^{-\rho} \frac{\left(\psi(k{-}\rho){-}{\rm ln}\frac{z}{2}\right)}{\Gamma(k-\rho+1)} -
\left(\frac{z}{2}\right)^{\rho} \frac{\left(\psi(k{+}\rho){+} {\rm ln} \frac{z}{2}\right)}{\Gamma(k+\rho+1)}
\right] \right\}.
\label{MD21}
\end{equation}
Here $\psi(s)$ is the logarithmic derivative of the Euler Gamma function \cite{Bateman}
\begin{equation}
\psi(s) \equiv \frac{\Gamma^{\prime}(s+1)}{\Gamma(s+1)}
\,.
\label{MD3}
\end{equation}
Then, we use the properties of this function:
$$
\psi(-s) = \psi(s-1) +  \Gamma(s)\Gamma(1-s) \cos{\pi s} \,,
$$
\begin{equation}
\psi(0) = - \gamma \,, \quad \psi(n>0) = - \gamma + 1 + ... + \frac{1}{n}
\,,
\label{MD4}
\end{equation}
where $\gamma $ is the Euler - Mascheroni constant
\begin{equation}
\gamma \equiv \lim_{n \to \infty} \left(\sum_{k=1}^{n}\frac{1}{k} - {\rm ln} n \right) \simeq 0.57722 \,,
\label{MD5}
\end{equation}
and find that two first terms in the decomposition with respect to small argument $z$ are
\begin{equation}
K_2(z) \to \frac{2}{z^2} - \frac12 + ... \,.
\label{MD6}
\end{equation}
Next terms are of the order $\propto z^2 {\rm ln}z$; asymptotically they tend to zero, so we restrict our-selves only by the decomposition (\ref{MD6}).
Using this approximation in the formula for the local charge density
\begin{equation}
{\cal I}_{*} \equiv {-}  \frac{N^{\frac32}(r^{*})}{N^2(r)}\sum_{\rm a} 4\pi e_{\rm a} {\cal N}_{\rm a}(r^{*})\left\{1 {+} \frac14 \left[\lambda^2(r_{*}){-} \lambda^2(r) \right] \right\}
\label{u3}
\end{equation}
we see that the main term disappears because of the electro-neutrality condition $\sum_{\rm a} e_{\rm a} {\cal N}_{\rm a}(r^{*})=0$. The term of the next order gives
$$
{\cal I}_{*}(r) =  \left[ \frac{N(r)- N(r_{*})}{N^2(r)}\right] {\cal Q}_{*} \,,
$$
\begin{equation}
{\cal Q}_{*} \equiv \sum_{\rm a} \pi e_{\rm a} {\cal N}_{\rm a}(r^{*})\lambda_{\rm a}^2(r_{*})N^{\frac12}(r^{*}) \,.
\label{u35}
\end{equation}
The screening radius is predetermined by the formula
\begin{equation}
\frac{1}{{\cal R}^2_{*}(r)} \simeq  \sum_{\rm a} \frac{4\pi e^2_{\rm a} {\cal N}_{\rm a}(r^{*}) N(r^{*})}{k_{B}T(r_{*})N^2(r)}
\equiv  \frac{1}{\lambda^{2}_{\rm D} N^2(r)} \,.
\label{u2}
\end{equation}
The corresponding key equation for the electric potential  is of the form
$$
\frac{1}{r^2}\frac{d}{dr} \left[r^2 \frac{d \Theta}{dr}  \right]
- \left[\frac{1}{\lambda^2_{D} N^2(r)} + \frac{\mu^2}{r^4 \Psi^2_0 N(r)} \right]\Theta =
$$
\begin{equation}
=\left[ \frac{N(r)- N(r_{*})}{N^2(r)}\right] {\cal Q}_{*} - \frac{\mu K}{r^4 N(r)}\,.
\label{u4}
\end{equation}
If this equation is solved and the profile $\Theta(r)$ is reconstructed, one can find the pseudoscalar field distribution using the formula (\ref{If2}). Let us analyze the solutions in the so-called far zone and near zone.

\subsubsection{Far zone}

In the far zone, where $\frac{r^2}{\sqrt{N(r)}}>> r_{\rm A} \lambda_{D}$ (or, in fact $r>>\sqrt{r_{\rm A} \lambda_{D}}$), the basic equation (\ref{u4}) takes the form
$$
\frac{d^2 \Theta}{dx^2} +  \left(\frac{2}{x}\right) \frac{d \Theta}{dx}  - \left(\frac{r^2_{+}}{\lambda^2_{D}}\right) \Theta = {\cal J}(x) \,,
$$
\begin{equation}
{\cal J}(x) \equiv  \left[\frac{N(x)- N(x_{*})}{N^2(x)}\right] {\cal Q}_{*} r^2_{+} \,.
\label{ep344}
\end{equation}
In this case, the plasma charge density is the main producer of the electric field, and the axionically induced electric field associated with the magnetic charge can be considered as vanishing.
The solution to (\ref{ep344}) can be presented in the following form:
$$
\Theta(x) = \frac{1}{x}  \left\{C^{*} \sinh{\left[\frac{r_{+}}{\lambda_{D}}(x-x_{*}) \right]} + \right.
$$
\begin{equation}
\left.
+\frac{\lambda_{D}}{r_{+}}\int_{x_{*}}^x zdz {\cal J}(z)\sinh{\left[\frac{r_{+}}{\lambda_{D}}(x-z) \right]} \right\} \,.
\label{ep777}
\end{equation}
Clearly, this function satisfies the initial condition $\Theta(x_{*})=0$ automatically.
The constant of integration has to be chosen as follows:
\begin{equation}
C^{*} = - \frac{\lambda_{D}}{r_{+}}\int_{x_{*}}^{\infty} zdz {\cal J}(z) \exp{\left[\frac{r_{+}}{\lambda_{D}}(x_{*}-z) \right]}  \,,
\label{ep78}
\end{equation}
if we assume, that the potential $\Theta(x)$ is characterized by the following asymptotical behavior:
\begin{equation}
\lim_{x\to \infty} \Theta(x) = const \,, \quad  \lim_{x\to \infty} \Theta^{\prime}(x) = 0 \,.
\label{ep79}
\end{equation}
Indeed, combining (\ref{ep777}) and (\ref{ep78}) we can rewrite the electric potential in the form
$$
\Theta {=} {-}\frac{\lambda_{D}}{x r_{+}}  \left\{\sinh{\left[\frac{r_{+}}{\lambda_{D}}(x{-}x_{*}) \right]} \int_x^{\infty}zdz {\cal J}(z) e^{\frac{r_{+}}{\lambda_{D}}(x_*{-}z)} {+}
\right.
$$
\begin{equation}
\left.
{+} e^{\frac{r_{+}}{\lambda_{D}}(x^*{-}x)} \int_{x_{*}}^x zdz {\cal J}(z)\sinh{\left[\frac{r_{+}}{\lambda_{D}}(z{-}x_*) \right]} \right\} \,,
\label{2ep777}
\end{equation}
and consider the limit $x \to \infty$. Using the l'Hospital rule for the first and second terms we obtain
\begin{equation}
\Theta(x \to \infty) \to  -\frac{\lambda^2_{D}}{ r^2_{+}}  {\cal J}(x \to \infty)\,.
\label{3ep777}
\end{equation}
In order to illustrate and motivate the choice of the constant of integration $C^{*}$, let us consider the very far zone, in which $N \to 1$, and the source ${\cal J}(x)$ in (\ref{ep344}) takes the form
\begin{equation}
{\cal J}(x) \simeq  \frac{{\cal Q}_{*} r^2_{+} r_g}{r_{*}} \left(\frac{x-x_{*}}{x} \right)\,.
\label{Ap1}
\end{equation}
The integral with this function can be calculated analytically yielding
$$
\Theta(x) = \frac{1}{x}  \sinh{\left[\frac{r_{+}}{\lambda_{D}}(x-x_{*})\right]} \left[C^{*} + \frac{\lambda^3_{D} r_g {\cal Q}_{*}}{r_{*} r_{+}}\right] +
$$
\begin{equation}
+
\left(\frac{x_{*}}{x}-1 \right) \frac{\lambda^2_{D} r_g {\cal Q}_{*}}{r_{*}} \,.
\label{Ap2}
\end{equation}
The growing branch of this solution can be eliminated if and only if
\begin{equation}
C^{*} = -  \frac{\lambda^3_{D} r_g {\cal Q}_{*}}{r_{*} r_{+}}\,.
\label{Ap3}
\end{equation}
Thus the solution for the electromagnetic potential in the very far zone is of the Coulombian type
\begin{equation}
\Theta(x) = \left(\frac{x_{*}}{x}{-}1 \right) \frac{\lambda^2_{D} r_g {\cal Q}_{*}}{r_{*}} \,,
\Theta^{\prime}(x) = {-} \frac{\lambda^2_{D} r_g {\cal Q}_{*}}{r_{+} x^2} \,.
\label{Ap4}
\end{equation}
Integration in (\ref{If2}) yields now
$$
\Phi(x)   = \Phi(x_*) + \frac{K}{r_{+}(1-a)} \ {\rm ln} \left|\frac{(x-1)(x_*-a)}{(x-a)(x_*-1)} \right| +
$$
\begin{equation}
+
\frac{r^2_{\rm A} \lambda^2_{D} r_g {\cal Q}_{*}}{\mu r_{+}r_{*} a (1{-}a)} {\rm ln} \left[\left(\frac{x{-}a}{x_* {-}a}\right)^{\gamma_1} \left(\frac{x{-}1}{x_* {-}1}\right)^{\gamma_2}\left(\frac{x}{x_*}\right)^{\gamma_3} \right],
\label{If27}
\end{equation}
where the following auxiliary parameters are introduced
$$
\gamma_1 = x_* - a > 0 \,, \quad  \gamma_2 = -a(x_* -1) <0 \,,
$$
\begin{equation}
\gamma_3 = -x_*(1-a)<0 \,, \quad \gamma_1 + \gamma_2 + \gamma_3 = 0 \,.
\label{If28}
\end{equation}
The asymptotic value of the axionic field is the constant equal to
$$
\Phi(\infty)   = \Phi(x_*) + \frac{K}{r_{+}(1-a)} \ {\rm ln} \left|\frac{(x_*-a)}{(x_*-1)} \right| -
$$
\begin{equation}
-
\frac{r^2_{\rm A} \lambda^2_{D} r_g {\cal Q}_{*}}{\mu r_{+}r_{*} a (1{-}a)} {\rm ln} \left[\left(x_* {-}a\right)^{\gamma_1} \left(x_* {-}1\right)^{\gamma_2} x_*{}^{\gamma_3} \right] \,.
\label{If24}
\end{equation}
Fig.3 contains illustrations of the profile of the function $\Phi$ as a function of the variable $x$ and of the guiding parameter $a$.
\begin{figure}
	\includegraphics[width=90mm,height=70mm]{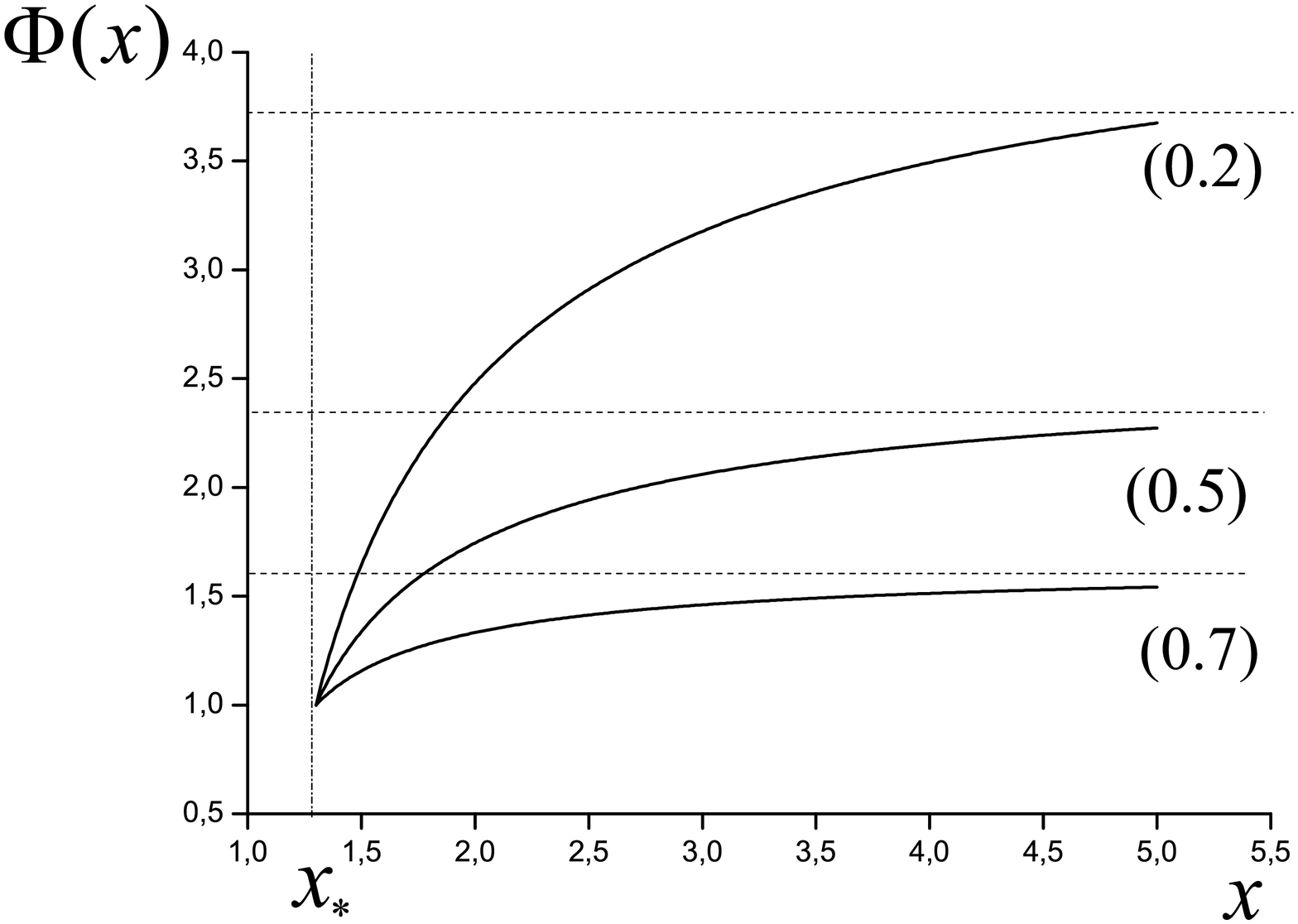}
	\caption{Plot of the function $\Phi(x,a)$ given by the formula (\ref{If27}). The values of the guiding parameter $a$ are indicated in parentheses near the corresponding curve. All curves have the same initial value $\Phi(x_*){=}1$; asymptotic values $\Phi(\infty)$ depend on the parameter $a$ and are marked by the corresponding horizontal lines. The coefficients in front of logarithmic functions in (\ref{If27}) are chosen equal to 1.5 and 2, respectively, for the sake of simplicity.}
\end{figure}

\subsubsection{Near zone}

In the near zone, where $\frac{r^2}{\sqrt{N(r)}}<< r_{\rm A} \lambda_{D}$, the term with axionic scale parameter $r_{\rm A}$ is the leading-order term;
in this limit we have to solve the equation
\begin{equation}
\frac{1}{r^2}\frac{d}{dr} \left(r^2 \frac{d \tilde{\Theta}}{dr} \right)  {-} \frac{r^2_{\rm A} \tilde{\Theta}}{r^4 N(r)} = \left[ \frac{N(r){-} N(r_{*})}{N^2(r)}\right] {\cal Q}_{*} \,,
\label{0ep337}
\end{equation}
where the function we search for is
\begin{equation}
\tilde{\Theta} \equiv \Theta - \frac{\mu K}{r^2_{\rm A}} \,.
\label{ep337}
\end{equation}
In terms of variable $x$, the equation (\ref{0ep337}) can be written as
\begin{equation}
\frac{d^2 \tilde{\Theta}}{dx^2} + \left(\frac{2}{x} \right) \frac{d \tilde{\Theta}}{dx}  - \frac{r^2_{\rm A}\tilde{\Theta}}{r^2_{+} x^2(x-1)(x-a)}  = {\tilde {\cal J}}(x) \,,
\label{NZ1}
\end{equation}
where the source term ${\tilde {\cal J}}$ is of the form
\begin{equation}
{\tilde {\cal J}} = {\cal Q}_{*} \ r^2_{+} \ \frac{x^2 \left[x^2\left(1-N(r_{*})\right) - x(a+1)+ a  \right]}{(x-1)^2 (x-a)^2}  \,.
\label{au1w}
\end{equation}
The linear equation (\ref{NZ1}) is an inhomogeneous version of the Fuchs-type equation (\ref{Fu1}) with
\begin{equation}
b{=}\alpha {=} \delta {=} \epsilon_1 {=} \epsilon_2 {=} p_1 = 0 \,, \quad \gamma {=} 2 \,, \quad p_2 {=} -\frac{r^2_{\rm A}}{r^2_{+}} \,.
\label{au49}
\end{equation}
Using the Fuchsian condition (\ref{Fu3})
we obtain additionally $\beta=1$. Formally speaking, when the fundamental solutions $Y_{(1)}(x)$ and $Y_{(2)}(x)$ to the homogeneous equation (\ref{NZ1}) are reconstructed, we obtain
\begin{equation}
\tilde{\Theta} = C_1 Y_{(1)} + C_2 Y_{(2)} + Y^{*}\,,
\label{a33}
\end{equation}
where $Y^{*}$ is the particular solution to the equation (\ref{NZ1}):
$$
Y^{*}(x) = \frac{1}{x^2_{*} W(x_{*})} \int_{x_{*}}^x z^2 dz {\tilde {\cal J}}(z) \times
$$
\begin{equation}
\times \left[Y_{(1)}(z) Y_{(2)}(x) - Y_{(1)}(x)Y_{(2)}(z)\right]\,.
\label{a34}
\end{equation}
The Wronsky determinant $W(x)$ is represented using the Liouville theorem as follows:
$$
W(x) = \frac{x^2_{*}}{x^2} W(x_{*}) \,,
$$
\begin{equation}
W(x_{*}) {=} \left[Y_{(1)}(x_{*}) Y^{\prime}_{(2)}(x_{*}) {-} Y^{\prime}_{(1)}(x_{*})Y_{(2)}(x_{*})\right]\,.
\label{a35}
\end{equation}
Typical sketches for numerical solutions to the equation (\ref{NZ1}) are presented in Fig.4.
\begin{figure}
	\includegraphics[width=75mm,height=70mm]{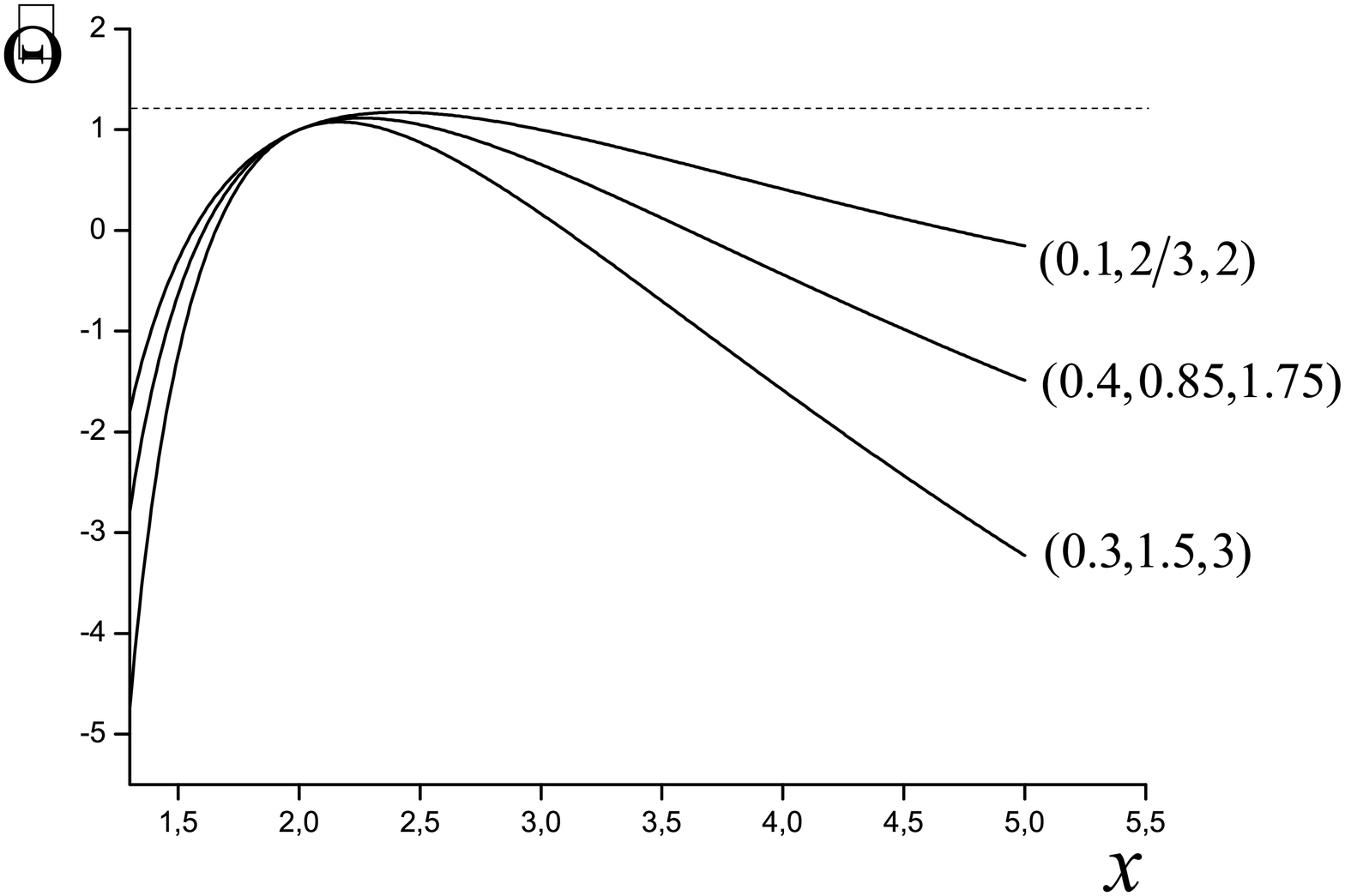}
	\caption{Plot of the function $\tilde{\Theta}$, which illustrates the profile of the reduced electric potential in the near zone. The curves are distinguished by three parameters indicated in parentheses near the corresponding curve; the first value relates to the parameter $a$, the second value describes the guiding parameter $\frac{r^2_{\rm A}}{r^2_{+}}$; the third value relates to the parameter $Q_{*}$. The general feature of these curves is the presence of maximum of the electric potential at $x=x_{\rm max}$ in the near zone; at $x<x_{\rm max}$ the electric field $E(x)= \frac{1}{r_{+}} \frac{d\tilde{\Theta}}{dx}$ is positive, at $x>x_{\rm max}$ it is negative. Since the electric field changes sign, we deal with the example of plasma stratification.}
\end{figure}

\subsection{Plasma with generalized Lorentz force, $\nu=1$}

In the regime $\left|\frac{e_{\rm a}\Psi}{k_{B}T_0}\right|<<1$, the key equation for the super-potential $\Psi$ (\ref{IIeq17}) transforms into
\begin{equation}
\frac{1}{r^2}\frac{d}{dr} \left(r^2 \frac{d \Psi}{dr} \right)  - \frac{1}{{\cal R}^2_{*}} \Psi =
{\cal I}_{*} \,.
\label{2plus1}
\end{equation}
Clearly, only the plasma source ${\cal I}_{*}$ and the plasma screening radius ${\cal R}_{*}$ predetermine the behavior of the super-potential; the axionic parameters do not enter the equation for the super-potential in contrast to the case $\nu=0$, and in this sense we deal with a kind of "pure relativistic Pannekoek-Rosseland" model. The results for non-relativistic and ultra-relativistic plasmas can now be extracted from the corresponding paragraphs of the previous Subsection. For instance, for the super-potential $\Psi$ one can use solutions obtained for $\Theta$ (\ref{u4}) with $\mu=0$, and then find the axion field with the source correspondingly reconstructed.

\section{Relativistic Boltzmann electron-positron plasma}

Let us consider the electron-positron plasma, for which $m_{\rm p}=m_{\rm e}=m$, $e_{\rm p}=-e_{\rm e}=e$ and ${\cal N}_{\rm p}= {\cal N}_{\rm e}={\cal N}$. When the axion field is absent, the equation for the electric field takes the form
$$
\frac{1}{r^2}\frac{d}{dr} \left(r^2 \frac{d \Theta}{dr}\right) =
$$
\begin{equation}
= \frac{8\pi e {\cal N}(r^{*})}{\sqrt{N(r)}}  \left[\frac{\lambda(r^{*})K_2(\lambda(r))}{\lambda(r)K_2(\lambda(r^{*}))}\right]  \sinh{\left(\frac{e \Theta}{k_{B}T_0}\right)} \,.
\label{EP1}
\end{equation}
Clearly, $\Theta(r) \equiv 0$ satisfies this equation, and the necessary condition $\Theta(r_*)=0$ does not contradict this solution. The trivial solution $\Theta(r) \equiv 0$ is unique for this equation of the second order in ordinary derivatives, if one of the following two requirements is satisfied. First, when $\Theta^{\prime}(r_*)=0$, we deal with the unique solution of the Cauchy problem with the electric potential and electric field vanishing on the sphere of the radius $r_*$. Second, when $\Theta(\infty)=0$, $\Theta^{\prime}(\infty)=0$, we deal with a special requirement based on ansatz of Coulombian behavior of the electric field in the asymptotical regime.
From physical point of view  the solution $\Theta(r)=0$ is well-motivated in this context, since there is no electric source in this case. In other words, the gravitationally induced Pannekoek-Rosseland  effect is absent in the electrically neutral electron-positron plasma without axionic environment of the magnetic monopole. Let us analyze now the situation with the axionic field interacting with magnetic field of the monopole and with the plasma.

\subsection{Electron-positron plasma with symmetric generalized Lorentz force, $\nu =1$}

We introduce again the super-potential (\ref{IIaxion92})
and obtain directly that it obeys the equation, which differs from (\ref{EP1}) only by the replacement $\Theta \to \Psi$:
$$
\frac{1}{r^2}\frac{d}{dr} \left(r^2 \frac{d \Psi}{dr}\right) =
$$
\begin{equation}
 =\frac{8\pi e {\cal N}(r^{*})}{\sqrt{N(r)}}  \left[\frac{\lambda(r^{*})K_2(\lambda(r))}{\lambda(r)K_2(\lambda(r^{*}))}\right]  \sinh{\left(\frac{e \Psi}{k_{B}T_0}\right)}
  \,.
\label{nul1}
\end{equation}
Again, there is the trivial solution to this equation, $\Psi(r) \equiv 0$, and the condition $\Psi(r_*)=0$ is satisfied by definition.  This trivial solution can be realized if, e.g., $\Psi^{\prime}(r_*)=0$, but this requirement, which connects the values of electric and axionic fields on the sphere $r=r_*$, is specific. In general, $\Psi^{\prime}(r_*) \neq 0$, and  the solution is not trivial. Both cases are interesting in our model; let us start with the first one.

\subsubsection{The model of self-compensation of the Lorentz force}

 When we consider the equation (\ref{nul1}) with conditions   $\Psi(r_*)=0$ and $\Psi^{\prime}(r_*)=0$, we obtain from the theorem of existence and uniqueness that the solution is trivial, i.e.,  $\Psi(r) \equiv 0$. In other words, the electric potential $\Theta(r)$ is not constant, the radial electric field $E(r)$ is not vanishing, but the profile of $E(r)$ is predetermined by the profile of the axion field. Indeed, when $\Psi(r) \equiv 0$, the electric potential $\Theta(r)$ is
$$
\Theta(r)  =  - \mu \int_{r^{*}}^r \frac{ dr \Phi(r)}{r^2} =
$$
\begin{equation}
= - \frac{\Psi^2_0}{\mu} \left[r^2 N(r) \frac{d\Phi}{dr} - r^2_{*} N(r_{*}) \frac{d\Phi}{dr}_{|r=r_{*}} \right] \,,
\label{nul2}
\end{equation}
and $\Phi(r)$ itself satisfies the equation:
\begin{equation}
r^2\frac{d}{dr} \left[r^2 N(r) \frac{d \Phi}{dr}\right] = \frac{\mu^2 \Phi}{\Psi^2_0}  \,.
\label{nul4}
\end{equation}
The solution to the last equation is already described (compare with (\ref{004})-(\ref{Fu31})). The radial electric field is presented by the formula
\begin{equation}
E(r) \equiv \sqrt{-F_{r0}F^{r0}} = \Theta^{\prime}(r) = - \frac{\mu}{r^2} \Phi(r)                                                                                                                       \,.
\label{nul5}
\end{equation}
Clearly, the electric field in plasma is nonvanishing; it is produced by the interaction between the magnetic field of the monopole and the axion field. Nevertheless there is no plasma polarization in this electric field. Why is this possible? The explanation is that in this case the generalized Lorentz force (\ref{kin3}) vanishes, since the standard electric field $F^r_{\ \ 0}$ is compensated by the axionically induced electric field $\phi F^{*r}_{ \ \ \ 0}$.

\subsubsection{On the properties of nontrivial solutions}

Now we consider the equation (\ref{nul1}) with conditions   $\Psi(r_*)=0$ and $\Psi^{\prime}(r_*) \neq 0$, and obtain non-trivial distribution of the super-potential $\Psi(r)$.
The master equation for the super-potential in the leading order high-temperature approximation takes the form
\begin{equation}
\frac{1}{r^2}\frac{d}{dr} \left(r^2 \frac{d \Psi}{dr}\right)
= \frac{\Psi}{\lambda^2_{D} N^2(r)} \,,
\label{nul6}
\end{equation}
\begin{equation}
\frac{1}{\lambda^2_{D}} =  \frac{8\pi e^2 {\cal N}(r^{*}) N^{\frac32}(r_*)}{k_{B}T_0}   \,,
\label{nul67}
\end{equation}
$\lambda_{D}$ is the screening parameter defined above.
In terms of dimensionless variable $x$ the master equation (\ref{nul6}) converts into
\begin{equation}
\frac{1}{x^2}\frac{d}{dx} \left(x^2 \frac{d \Psi}{dx}\right)
=\Psi \left(\frac{r^2_{+}}{\lambda^2_{D}}\right) \left[ \frac{x^2}{(x{-}1) (x{-}a)} \right]^2 \,.
\label{nul8}
\end{equation}
If we consider the case, where the object boundary is far from the horizons, i.e.,  $x_{*} >> 1 > a$, the solution to this equation with the condition $\Psi(x_{*})=0$ can be modeled as follows
\begin{equation}
\Psi(x) = \Psi^{\prime}(x_*) \left(\frac{x_*}{x}\right) \left(\frac{\lambda_{D}}{r_{+}}\right) \sinh{\left[\frac{r_{+}}{\lambda_{D}}(x{-}x_*)\right]}.
\label{nul9}
\end{equation}
This solution takes infinite value at $x \to \infty$. Numerical analysis of the equation (\ref{nul8}) confirms this statement. In other words, there are no physically motivated solutions with finite asymptotes, thus it is reasonable to consider only the trivial solution $\Psi(x) \equiv 0$.

\subsection{Electron-positron plasma with $\nu=0$}

If the Lorentz force is not modified by the axion-plasma coupling, i.e., $\nu=0$, searching for the electric potential we deal with the following decoupled master equation:
$$
\frac{1}{r^2}\frac{d}{dr} \left(r^2 \frac{d \Theta}{dr}\right)  - \frac{\mu^2 \Theta}{r^4 N(r)\Psi^2_0}
-
$$
\begin{equation}
- \frac{8\pi e {\cal N}(r^{*})}{\sqrt{N(r)}}  \left[\frac{\lambda(r^{*})K_2(\lambda(r))}{\lambda(r)K_2(\lambda(r^{*}))}\right] \sinh{\frac{e \Theta}{k_{B}T_0}}
=-  \frac{\mu K }{r^4 N(r)}  \,.
\label{IIIeq17}
\end{equation}
The pseudoscalar field itself can be found from the equation
\begin{equation}
N r^2 \frac{d \Phi}{dr}   = K - \frac{\mu}{\Psi^2_0} \Theta  \,,
\label{IIIaxion92}
\end{equation}
when the potential $\Theta$ is obtained. In the high-temperature electron-positron plasma, i.e., when
$\left|\frac{e \Theta}{k_{B}T_0} \right|<<1$, we obtain the reduced equation:
\begin{equation}
\frac{1}{r^2}\frac{d}{dr} \left(r^2 \frac{d}{dr}\Theta \right)  -   \frac{\Theta}{\Re^2}   = -  \frac{\mu K }{r^4 N(r)} \,,
\label{ep3}
\end{equation}
where we introduce the effective screening radius $\Re$ by the definition
\begin{equation}
\frac{1}{\Re^2} \equiv  \frac{\mu^2}{r^4 N(r)\Psi^2_0} {+}  \frac{8\pi e^2 {\cal N}(r^{*})}{k_B T_0 \sqrt{N(r)}}  \left[\frac{\lambda(r^{*})K_2(\lambda(r))}{\lambda(r)K_2(\lambda(r^{*}))}\right]   \,.
\label{ep4}
\end{equation}
In the ultra-relativistic limit we deal, respectively, with the formula
\begin{equation}
\frac{1}{\Re^2} \equiv  \frac{\mu^2}{r^4 N(r)\Psi^2_0} +  \frac{1}{\lambda^2_{D} N^2(r)}  \,.
\label{ep5}
\end{equation}
On the right-hand side of the equation (\ref{ep3}) there is only one source term; it relates to the electric field produced by the axion field in the presence of magnetic charge of the monopole.
This source guarantees the plasma to be polarized in contrast to the case with $\nu=1$, the plasma polarization being essential in the far zone. In order to discuss new features of the electric field profile, let us focus on the case $r>r_{*}>>r_{+}$.
Now we deal with the key equation
\begin{equation}
\frac{1}{x^2}\frac{d}{dx} \left(x^2 \frac{d \Theta}{dx} \right)  {-}   \frac{r^2_{+}}{\lambda^2_{D}} \Theta   = -  \frac{\mu K }{r^2_{+} x^2 (x{-}1)(x{-}a)} \,.
\label{ep37}
\end{equation}
Using the formula (\ref{ep777}) we obtain now
$$
\Theta(x) = \frac{\mu K \lambda_{D}}{x r_{+}}  \left\{C^{**} \sinh{\left[\frac{r_{+}}{\lambda_{D}}(x-x_{*}) \right]} -
\right.
$$
\begin{equation}
\left.
- \int_{x_{*}}^x \frac{dz}{z(z-1)(z-a)} \sinh{\left[\frac{r_{+}}{\lambda_{D}}(x-z) \right]} \right\} \,,
\label{ep888}
\end{equation}
where the constant of integration is chosen as
\begin{equation}
C^{**} = \int_{x_{*}}^{\infty} \frac{dz}{z (z-1)(z-a)} \exp{\left[\frac{r_{+}}{\lambda_{D}}(x_{*}-z) \right]}  \,,
\label{ep789}
\end{equation}
providing the electric potential to tend to finite value at infinity.
The electric field is given by the formula
$$
E(x) \equiv \frac{1}{r_{+}}\Theta^{\prime}(x) =
$$
$$
- \frac{\Theta(x)}{x r_{+}} + \frac{\mu K }{x r_{+}}  \left\{C^{**} \cosh{\left[\frac{r_{+}}{\lambda_{D}}(x-x_{*}) \right]} -
\right.
$$
\begin{equation}
\left.
- \int_{x_{*}}^x \frac{dz}{z(z-1)(z-a)} \cosh{\left[\frac{r_{+}}{\lambda_{D}}(x-z) \right]} \right\} \,.
\label{ep001}
\end{equation}
The value of this electric field at $r=r_{*}$ is
\begin{equation}
E(r_{*}) = \frac{\mu K}{r_{*}} \int_{x_{*}}^{\infty} \frac{dz}{z (z{-}1)(z{-}a)} \exp{\left[\frac{r_{+}}{\lambda_{D}}(x_{*}{-}z) \right]} \,.
\label{ep71}
\end{equation}
At infinity, using the asymptotic formula (\ref{3ep777}) we obtain
\begin{equation}
\Theta(x) \to  \mu K \frac{\lambda^2_{D}}{ r^4_{+} x^4} \,, \quad E(x) \to \  - 4 \mu K \frac{\lambda^2_{D}}{ r^5_{+} x^5}  \,.
\label{7ep777}
\end{equation}
Since $\Theta(r_*) = 0 = \Theta(\infty)$, there is an extremum in the profile of the function $\Theta(r)$. If to indicate the radius, corresponding to the extremum,  as $r_{**}$, we see that the  radial electric field $E(r)= \Theta^{\prime}(r)$ takes zero value at this radius, i.e., $E(r_{**})=0$. Depending on the sign of the parameter $\mu K$, we obtain positive or negative value of the electric field on the basic sphere $r=r_{*}$. There are two possible variants of behavior of the electric field profile.
First, when $\mu K<0$, the value $E(r_{*})$ is negative; then the function $E(r)$ grows, reaches zero value at $r=r_{**}$, changes sign, reaches the maximum at $r=r_{\rm max}$, and then tends asymptotically to zero.
Second, when $\mu K>0$, the value $E(r_{*})$ is positive; then the function $E(r)$ tends to zero, changes the sign, reaches the minimum at $r=r_{\rm min}$, and then tends asymptotically to zero. In both cases, in the zone $r_{*}<r<r_{\rm extremum}$ the electric field changes sign, so this domain in the electron-positron plasma is polarized and demonstrates the effect of stratification.

\section{Degenerated electron plasma}

Let us consider the model of degenerated electron gas with $T_0=0$ and with standard Lorentz force ($\nu {=}0$); and let it be the non-relativistic gas, i.e., ${\cal P}_{\rm F}(r_{*}) << m_{\rm e} c$.
Now we have to analyze the following reduced equation for the electric potential:
$$
\frac{1}{r^2}\frac{d}{dr} \left(r^2 \frac{d \tilde{\Theta}}{dr} \right) - \frac{r^2_{\rm A}}{N(r) r^4} \tilde{\Theta} =
$$
\begin{equation}
= {-} \frac{32 e \pi^2 m^3_{\rm e}c^3}{h^3 N^2(r)} \left[N(r_{\rm F}) -  N(r)\right]^{\frac32}
 \,.
\label{FD29}
\end{equation}
where, again, $\tilde{\Theta}= \Theta -\frac{\mu K}{r^2_{\rm A}}$. In terms of the variable $x$
this equation has the same form as (\ref{NZ1}), however, the source ${\tilde {\cal J}}$ is now equal to
\begin{equation}
{\tilde {\cal J}} = {-} \frac{32 e \pi^2 m^3_{\rm e} c^3  r^2_{+} x}{h^3 (x{-}1)^2 (x{-}a)^2} \left[x^2\left(N(r_{\rm F}){-}1\right) {+} x(a{+}1){-} a  \right]^{\frac32}.
\label{DG33}
\end{equation}
The analysis of the solution is similar to the previous case: again it is the solution of the inhomogeneous version of the Fuchs-type equation (\ref{Fu1}); only one detail has to be taken into account, that now the reduced radial variable $x$ changes in the interval
$\frac{r_{*}}{r_{+}}<x<\frac{r_{\rm F}}{r_{+}}$, where the threshold radius $r_{\rm F}$ can be found from the equation (\ref{FD71}). In the zone $x>\frac{r_{\rm F}}{r_{+}}$  we have to solve homogeneous equation with ${\tilde {\cal J}}=0$, and see that asymptotically the electric field is Coulombian: $\Theta^{\prime} \propto \frac{1}{r^2}$.

\section{Discussion}

Spectral characteristics of astrophysical objects are sensible to the presence or absence  of electric fields in their environs. Astrophysicists search for fine details of the object spectra as fingerprints of electric fields.
Theoreticians elaborate simple, complex and sophisticated models for the interpretation of peculiarities of these spectra. We work as theoreticians with the
model of axionic magnetic monopole (dyon), which became a very interesting "theoretical laboratory" after the work \cite{Wilczek87}. Indeed, this model includes static spherically symmetric gravitational, magnetic, pseudoscalar (axion) fields, as well as, an axionically induced electric field, which interact with one another. We add the relativistic plasma into this model, and consider in detail the plasma polarization, stratification and the compensating electric field generated in plasma in the axionic and gravitational environment.

We consider the relativistic plasma as an example of axionically active system. What is the sense of the term {\it axionically active}? We trace three schemes of reaction of the plasma particles
on the action of pseudoscalar (axion) field associated with axionic dark matter. The first scheme describes an indirect interaction of the macroscopic type: the collective electromagnetic field in plasma is influenced by the axion field; the interaction of this type is analyzed in the framework of axion electrodynamics. The second scheme describes the direct microscopic interaction of plasma particles with axion field; this interaction is modeled by the generalized Lorentz force (\ref{kin3}), into which the term $\phi F^{*}_{ik}$ is introduced in addition to the standard Maxwell tensor $F_{ik}$. The third scheme is connected with the influence of the background strong gravitational field: we study the axionic analog of the Pannekoek-Rosseland effect of plasma polarization and stratification.

The difference between polarization and stratification in plasma can be visualized using the profile of the compensating electric field. We deal with the effect of polarization, when the profile of the electric potential $A_0(r)$ is monotonic, and the electric field $E(r)= A_0^{\prime}(r)$ does not change sign. The effect of plasma stratification corresponds to the case, when the electric potential has at least one extremum, so that the electric field takes zero values at some radii $r_{\rm s}$, i.e., $E(r_{\rm s}){=}0$, then changes sign (see Fig. 4). The exact results obtained in the work, show that both effects: polarization and stratification can be visualized in the profiles near the magnetic monopole with axionic environment (the axionic dyon).  How can one distinguish physical conditions, which guarantee the presence of stratification, and when only the plasma polarization exists? The answer is connected with the sophisticated interplay between guiding and structure parameters of the model.

In the model under discussion there are three guiding parameters and one structure parameter with the dimensionality of length (in other words, there are four scale parameters). The first guiding parameter, $K$, appears in (\ref{axion921}) as a constant of integration; in fact, it plays the role of effective axionic charge (see, e.g., \cite{Y,BZ17} for details). The second guiding parameter, $r_{\rm A}=\frac{\mu}{\Psi_0}$, describes the scale of an axionically induced conversion of the magnetic field into the electric. The third one, $\lambda_{D}$, characterizes the radius of screening  in the Boltzmann plasma; an analog of this parameter in the degenerated plasma with zero temperature is the Fermi radius, $r_{\rm F}$ (see (\ref{FD71})), at which the electron gas number density vanishes. The structure parameter, $r_{+}$, (\ref{2eq43}), describes the radius of an outer horizon of the dyon with the Reissner-Nordstr\"om metric.

The guiding parameters predetermine the properties of solutions of the key set of two coupled master equations, to which the total system of axionic, electromagnetic and kinetic equations is effectively reduced. We mean the equation of axion electrostatics (see (\ref{eq17}) for the Boltzmann plasma, and (\ref{FD27}) for the degenerated electron gas), and equation of the axion field (\ref{axion921}). The procedure of decoupling these two key master equations presents us the necessity to analyze the Fuchs-type equations (\ref{Fu1}). Fortunately, equations of this type attracted the attention of mathematicians from the 1920s; many properties of these solutions, which can be indicated as special functions of mathematical physics,  are well-documented in literature. We quote the mathematical works, in which these functions are studied, and, in addition, we use numerical calculations and typical plots for the illustration of main conclusions (see Figs. 1-4).

The model under consideration describes self-consistently the interaction of the axion field with the electric field and plasma. This means that the model describes both the influence of axion on the electromagnetic field of the dyon, and the backreaction of the dyon electric field on the state of the axion field. The first typical illustration of such mutual influence is given by the formula  (\ref{If1}). The profile of the axion field $\Phi(r)$ (\ref{If2}) is predetermined by the integral containing the electric potential, the guiding parameter being the ratio $\frac{\mu}{\Psi^2_0}$. There are estimations of the quantity $\frac{1}{\Psi_0}=g_{{\rm A}\gamma \gamma}<1.47 \cdot 10^{-10} {\rm GeV}^{-1}$ connected with the constant of the axion-photon coupling \cite{CAST14}, however, there is no adequate estimation of the magnetic charge $\mu$. As for the second typical illustration describing the backreaction of the plasma, one can see, e.g., from (\ref{If27})  that the profile of the axion field is regulated by the plasma parameter ${\cal Q}_{*}$ (\ref{u35}) and by the screening Debye parameter $\lambda_{D}$. Detailed estimations of these contributions are interesting, but are out of scope of this paper.

\vspace{3mm}

What are the main conclusions?

1. When we deal with relativistic Boltzmann electron - ion plasma, the plasma polarization and  electric field near the axionic dyon can be formed as follows. First, due to the Pannekoek - Rosseland effect, induced under the influence of the strong gravitational field, the  isothermal plasma becomes polarized, and the compensating radial electric field appears. This collective electric field in plasma is influenced by the axion field on both: the microscopic and macroscopic levels (via the electrostatics equations and via the generalized Lorentz force, respectively). Second, the radial electric field generated by the magnetic field of the monopole in the axionic environment, additionally polarizes or depolarizes the plasma, depending on the sign of the effective axion charge; if the Pannekoek-Rosseland electric field and this axionically induced field act against each other, it can lead to plasma stratification in the near zone of the dyon. Third, the electric field is screened by the polarized plasma, thus providing the last structure element in the charged particle density near the dyon. Thus, {\it the first conclusion} is that in the relativistic Boltzmann electron - ion plasma three principal plasma configurations and three electric field profiles can be formed, which are indicated as follows: first, the simple plasma polarization, where the axionically induced electric field is directed along the Pannekoek-Rosseland field; second, the plasma stratification, where the Pannekoek-Rosseland electric field is directed contrarily to the axionically induced electric field; third, the profile related to the compensation of two mentioned electric fields.

2. When we deal with relativistic Boltzmann electron - positron plasma, the Pannekoek-Rosseland effect is absent because of the equivalence of the masses of electrons and positrons. Thus, the scheme of formation of the electric field is simpler than the one described above. Nevertheless, the axionically induced electric field again polarizes the plasma, and plasma back-reaction and screening form the final sophisticated profile of the coupled electric and axionic fields. {\it The second conclusion} is that in the relativistic Boltzmann electron - positron plasma  one can also find the plasma polarization, stratification and the self-compensation of the electric field.

3. When we consider the relativistic degenerated electron gas with zero temperature, we deal with the following scheme of formation of the electric field. First, the gravitational field of the dyon influences the Fermi momentum of electrons, ${\cal P}_{\rm F}(r)$, thus establishing the profile of the local number density ${\cal N}_{\rm e}(r)$. The inhomogeneity of this profile produces the electric field (which is generated, in fact, due to the gravity field influence). The axionically induced electric field, appears due to the transformation of the magnetic field of the monopole, again can be directed along and contrarily with respect to the electric field in the electron plasma, depending on the sign of the effective axionic charge. In contrast to the cases with Boltzmann plasma distributed in the infinite range $r_0<r<\infty$, the degenerated electron gas occupies the finite zone $r_0<r<r_{\rm F}$, the boundary of which is predetermined by the equation ${\cal P}_{\rm F}(r_{\rm F})=0$ (see (\ref{FD71})). {\it The third conclusion} is that both polarization and stratification can be found in the degenerated electron gas in the vicinity of axionic dyon.

4. In a special work, we hope to make estimations, to formulate some observational prognoses and apply these theoretical predictions to the real astrophysical sources with strong magnetic field.

\acknowledgments{Authors are grateful to Prof. Ignat'ev Yu.G. and Dr. Zayats A.E. for fruitful discussions. The work was supported by Russian Science Foundation (Project No. 16-12-10401), and by the Program of Competitive Growth of Kazan Federal University.}

\end{document}